\documentclass[twocolumn]{aastex631}

\usepackage{gensymb}
\usepackage{amsmath,amssymb,amsthm,amsfonts}
\usepackage{afterpage}
\usepackage{comment}
\usepackage{tabularx}
\usepackage{booktabs}
\usepackage[flushleft]{threeparttable}
\usepackage{nicefrac}

\accepted{March 8, 2024}

\begin{document}

\title{Investigating model dependencies for obscured Active Galactic Nuclei: a case study of NGC\,3982}

\correspondingauthor{Krist\'{i}na~Kallov\'{a}}
\email{kristina.kallova@mail.udp.cl}

\author[0009-0008-8860-0372]{Krist\'{i}na~Kallov\'{a}}
\affiliation{Instituto de Estudios Astrof\'isicos, Facultad de Ingenier\'ia y Ciencias, Universidad Diego Portales, Av. Ej\'ercito Libertador 441, Santiago, Chile\\}
\affiliation{Department of Theoretical Physics and Astrophysics, Faculty of Science, Masaryk University, Kotl\'{a}\v{r}sk\'{a} 267/2, 611 37 Brno, Czech Republic}

\author[0000-0001-9379-4716]{Peter~G.~Boorman}
\affiliation{Caltech, 1216 E. California Blvd., Pasadena, California 91107, USA}
\affiliation{Astronomical Institute of the Czech Academy of Sciences, Bo\v{c}n\'{i} II 1401, 141 00 Prague 4, Czech Republic}

\author[0000-0001-5231-2645]{Claudio~Ricci}
\affiliation{Instituto de Estudios Astrof\'isicos, Facultad de Ingenier\'ia y Ciencias, Universidad Diego Portales, Av. Ej\'ercito Libertador 441, Santiago, Chile\\}
\affiliation{Kavli Institute for Astronomy and Astrophysics, Peking University, Beijing 100871, People's Republic of China}

\begin{abstract}
X-ray spectroscopy of heavily obscured Active Galactic Nuclei (AGN) offers a unique opportunity to study the circum-nuclear environment of accreting supermassive black holes (SMBHs). However, individual models describing the obscurer have unique parameter spaces that give distinct parameter posterior distributions when fit to the same data. To assess the impact of model-specific parameter dependencies, we present a case study of the nearby heavily obscured low-luminosity AGN NGC\,3982, which has a variety of column density estimations reported in the literature. We fit the same broadband \textit{XMM-Newton}\,$+$\,\textit{NuSTAR} spectra of the source with five unique obscuration models and generate posterior parameter distributions for each. By using global parameter exploration, we traverse the full prior-defined parameter space to accurately reproduce complex posterior shapes and inter-parameter degeneracies. The unique model posteriors for the line-of-sight column density are broadly consistent, predicting Compton-thick $N_{\rm H}$\,$>1.5\times10^{24}\,\rm cm^{-2}$ at the 3$\sigma$ confidence level. The posterior median intrinsic X-ray luminosity in the 2--10\,keV band however was found to differ substantially, with values in the range log\,$L_{2-10\,{\rm keV}}$\,/\,erg\,s$^{-1}$\,=\,40.9--42.1 for the individual models. We additionally show that the posterior distributions for each model occupy unique regions of their respective multi-dimensional parameters spaces, and how such differences can propagate into the inferred properties of the central engine. We conclude by showcasing the improvement in parameter inference attainable with the \textit{High Energy X-ray Probe} (\textit{HEX-P}) with a uniquely broad simultaneous and high-sensitivity bandpass of 0.2--80\,keV. \\
\end{abstract}


\section{Introduction}
\label{sec:intro}

X-ray surveys have revealed that obscured Active Galactic Nuclei (AGN) with equivalent hydrogen column densities $N_{\rm H}$\,$>$\,10$^{22}$\,cm$^{-2}$ greatly outnumber AGN with less obscured sight lines, in the local Universe (e.g. \citealt{Ricci2017d}) as well as at higher redshifts (e.g., \citealt{Ueda2014,Buchner2014,Brandt2015}). Although in some cases obscuration could be produced by material in the host galaxy (e.g., \citealt{Buchner2017a,Gilli2022,Andonie2023}), the highest {\em Compton-thick} column densities ($N_{\rm H}\,>\,1.5\times10^{24}\,\rm cm^{-2}$) observed at low redshifts are likely dominated by a parsec-scale circum-nuclear obscurer surrounding the central accreting supermassive black hole (SMBH), akin to the axis-symmetric obscurer invoked in early unified schemes \citep{Antonucci1993,Urry1995,Netzer15,RamosAlmeida17}.

In the X-ray band, the combined effects of photo-electric absorption, fluorescence and Compton scattering from obscuring material give rise to a characteristic \lq reflection spectrum\rq\ that can dominate over any other AGN signatures if the line-of-sight $N_{\rm H}$ is sufficiently above the Compton-thick limit (i.e., $N_{\rm H}$\,$\gtrsim$\,5$\times10^{24}$\,cm$^{-2}$; \citealt{Setti1989,Murphy2009,Brightman2015}). Their prominent reflection spectrum thus makes X-ray spectroscopy of so-called \lq reflection dominated\rq\ Compton-thick AGN an ideal approach for probing the circum-nuclear environment of growing SMBHs. However, to infer useful information first requires a sufficiently sensitive broadband X-ray spectrum to be measured, that includes the flat underlying reflection continuum at $E$\,$\lesssim$\,10\,keV, the iron K$\alpha$ fluorescence line at 6.4\,keV and the Compton hump peaking at $\sim$~30\,keV \citep{Matt2000}. Secondly, inference is limited by the requirements of a physically-motivated model for the obscurer that must {\em uniquely} reproduce the broadband observed reflection spectrum from a combination of distinct model parameters.

The Nuclear Spectroscopic Telescope ARray (\textit{NuSTAR}; \citealt{Harrison2013}) is the first and currently-only focusing X-ray telescope in orbit capable of providing high-sensitivity spectroscopy $>$\,10\,keV. As a result, the combination of \textit{NuSTAR} ($E$\,$\sim$\,3--78\,keV) with sensitive soft X-ray facilities such as \textit{XMM-Newton} \citep{Jansen01}, \textit{Chandra} \citep{Weisskopf00}, \textit{Suzaku}/XIS \citep{Koyama07} and \textit{Swift}/XRT \citep{Burrows05} has been fundamental in measuring the broadband reflection spectra of numerous known (e.g., \citealt{Arevalo2014,Puccetti2014,Bauer2015,Annuar2015,Puccetti2016,Gandhi2017}) as well as previously unknown (e.g., \citealt{Gandhi2014,Boorman2016,Annuar2017,Sartori2018,Annuar2020,Kammoun2020}) Compton-thick AGN.

In terms of modelling, the earliest X-ray obscuration-based reflection models assumed semi-infinite plane geometries, and were originally designed to parameterise reflection from an accretion disc \citep{Magdziarz1995}. An improvement was provided by physically-motivated models featuring geometrically-thick obscurers in a specific geometry and finite optical depth (e.g., \citealt{Awaki1991}). Since the obscurer often has many geometric degrees of freedom, the reprocessed X-ray spectrum for a given model configuration cannot be determined analytically. It has hence become commonplace to produce spectral models via Monte Carlo Radiative Transfer methods in which X-ray photons are propagated through geometries of gas for a number of different parameter values describing the properties of the intrinsic AGN emission as well as the geometry and structure of the obscurer. Table models can then be created which feature a multi-dimensional discrete grid of parameters, with each parameter combination corresponding to a distinct spectrum. X-ray spectra are then fit to data via grid interpolation to solve for parameter values that optimise some fit statistic and reproduce the observed spectral data\footnote{For more information, see: \href{https://heasarc.gsfc.nasa.gov/docs/heasarc/caldb/docs/memos/ogip_92_009/ogip_92_009.pdf}{https://heasarc.gsfc.nasa.gov/docs/heasarc/caldb/docs/memos/ogip\_92\_009/ogip\_92\_009.pdf}}.

Recent years have seen a surge in not only different physically-motivated X-ray reflection table models for a variety of geometries \citep{Murphy2009,Brightman2011a,Balokovic2018,Tanimoto2019,Buchner2021,Ricci2023a}, but also bespoke packages designed to perform Monte Carlo Radiative Transfer simulations enabling arbitrary user-defined obscurer geometries and parameters (e.g., RefleX; \citealt{Paltani2017}, XARS; \citealt{Buchner2019}, SKIRT; \citealt{VanderMeulen2023}). However, with a plethora of publicly-available model geometries and simulation packages, models can often provide non-unique solutions when fitting observed X-ray spectral data (e.g., \citealt{Saha2022}). Non-unique solutions can arise on multiple levels, for example from the choice of parameter grids in a given table model, to wide-scale degeneracies between the parameters used to describe the AGN intrinsic spectrum as well as the surrounding obscurer. The issue is naturally more significant for fainter sources in which the observed reflection spectrum can be reproduced by a wider range of non-unique spectral shapes, which can then potentially affect the inference of parameters substantially from one model to the next.

A number of current X-ray obscurer models allow the user to add extra degrees of freedom to the modelling by decoupling the Compton-scattered continuum and/or fluorescence emission spectrum from the transmitted component, or by decoupling the line-of-sight column density from the global one. These decoupled models are often capable of describing more complex obscurer geometries than their default coupled configurations (see e.g., \citealt{Yaqoob2012}). For the highest signal-to-noise ratio data, multiple reflectors are often needed to reproduce the complex shapes of the underlying Compton-scattered continuum and Compton hump (e.g. Circinus galaxy and NGC~1068, \citealt{Arevalo2014, Bauer2015, Andonie2022}).

In this paper, we test the effect of non-unique spectral prescriptions for the obscurer with a case study of the Low-Luminosity ($L_\mathrm{2-10\,keV} \lesssim 10^{42}\,\rm{erg\,s^{-1}}$) Compton-thick AGN candidate NGC\,3982, a Seyfert~2 AGN located at a Hubble distance $18.91\pm1.33$\,Mpc\footnote{\href{https://ned.ipac.caltech.edu}{https://ned.ipac.caltech.edu}} (z\,=\,0.00371; \citealt{Martinsson2013}). Whilst the effects of model-dependencies will undoubtedly be larger for lower signal-to-noise spectra associated with higher-redshift targets (see e.g., \citealt{Buchner2014}), we chose NGC\,3982 because unlike the other handful of Compton-thick AGN known in a similar volume \citep{Asmus2020,Boorman2023}, NGC\,3982 has had a wide variety of column density estimations reported in the literature. \citet{Kammoun2020} fit broadband \textit{XMM-Newton}\,$+$\,\textit{NuSTAR} spectra of the source with \texttt{pexmon}, \texttt{MYTorus-coupled} and \texttt{MYTorus-decoupled}, finding Compton-thin as well as Compton-thick $N_{\rm H}$ values of $N_{\rm H}$\,$\sim$\,$0.48-4.5\,\times 10^{24}\,\mathrm{cm^{-2}}$. \citet{Saade2022} then fit the broadband \textit{Chandra}\,$+$\,\textit{NuSTAR} spectra with \texttt{borus02}, yielding an even higher Compton-thick line-of-sight column density of $N_{\rm H}\,>\,2\times10^{25}\,\rm{cm^{-2}}$. The large range of reported column density measurements in the literature suggests the source is of sufficient spectral sensitivity to test the effects of model-specific dependencies in constraining the properties of the circum-nuclear obscurer.

The structure of the paper is as follows: in \S\ref{sec:data} we summarize the X-ray data extraction, before describing our X-ray spectral fitting methodology in \S\ref{sec:method}. The key parameter posteriors together with luminosity and Eddington ratio inference are presented in \S\ref{sec:results} while discussion of the model-dependent degeneracies is presented in \S\ref{sec:discussion} together with the prospects attainable with the High Energy X-ray Probe \citep{Madsen2019}. We summarise our key findings in \S\ref{sec:summary}. For our luminosity calculations we assume the cosmological parameters from \citet{Planck2014}; $H_0 = 67.8\,\rm km\,s^{-1}\,Mpc^{-1}$, $\Omega_{\rm m} = 0.308$ and $\Omega_{\Lambda} = 0.692$.

\section{X-ray data}\label{sec:data}

\subsection{XMM-Newton}
The \textit{XMM-Newton} observation of NGC\,3982 was carried out on June 15th 2004. We obtained the archival data from the \textit{XMM-Newton} data archive\footnote{\href{http://nxsa.esac.esa.int/nxsa-web}{http://nxsa.esac.esa.int/nxsa-web}} (obs.\,ID:\,\,0204651201; PI:\,\,I.\,George). The Science Analysis Software (SAS; \citealp{Gabriel2004}) package was used to reprocess the raw Observation Data Files for all three cameras onboard \textit{XMM-Newton} (MOS1, MOS2 \& PN; \citealt{Struder01}) and to generate calibrated and concatenated EPIC event lists. The EPIC event lists were then filtered for flaring particle background via visual inspection of light curves in energy regions recommended by the SAS threads. Net exposure times after filtering accounted for 11.35\,ks for both MOS detectors and 9.14\,ks for PN. Source\,$+$\,background and background regions with radii of 49 and 65~arcsec, respectively, were then created using the corresponding EMOS camera images before extracting spectra with \texttt{evselect}. For EPN, the source\,$+$\,background and background regions were reduced to 45 and 50~arcsec, respectively, due to the central readout node proximity. Spectral response and effective area files were created using the \texttt{rmfgen} and \texttt{arfgen} commands, respectively.

\subsection{NuSTAR}
The \textit{NuSTAR} telescope observed NGC\,3982 on December 6th 2017. The data were downloaded from the HEASARC database\footnote{\href{https://heasarc.gsfc.nasa.gov/docs/archive.html}{https://heasarc.gsfc.nasa.gov/docs/archive.html}.} (obs.\,\,ID:\,60375001002; PI:\,\,M.\,Malkan) and processed for both Focal Plane Modules (FPMA \& FPMB) with the \textit{NuSTAR} Data Analysis Software (NuSTARDAS). The net exposure times of the observations for FPMA and FPMB were 33.41\,ks and 33.34\,ks, respectively. The task \texttt{nupipeline} was used to produce cleaned event files, before source\,$+$\,background circular regions with radii of 49~arcsec were created to encompass the source, making sure to account for any astrometric offsets by eye. Background circular regions with radii of 150~arcsec were then created to cover a large source-free part of the same detector as the source for each FPM. 

\begin{figure*}
    \includegraphics[width=\textwidth]{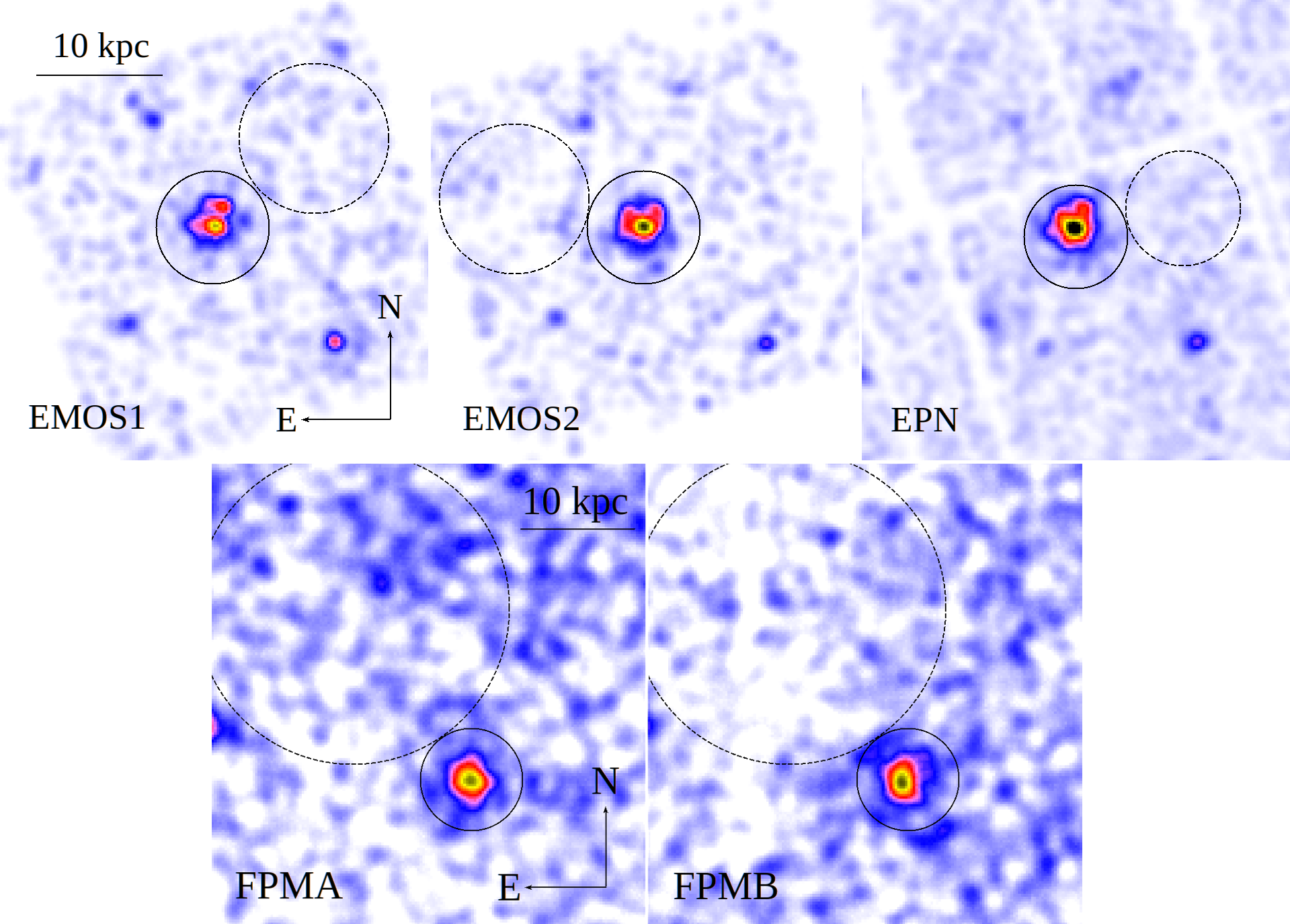}
    \caption{The $0.3-10.0$~keV \textit{XMM-Newton} (\textit{top}) and $3-78$~keV \textit{NuSTAR} (\textit{bottom}) X-ray images of NGC\,3982 showing the extraction regions for source\,$+$\,background (\textit{solid circles}) and background (\textit{dashed circles}) spectra. The source\,$+$\,background regions have radii of 49~arcsec for the EMOS and FPM detectors, and 45~arcsec for EPN due to the central readout node. The background regions have radii of 65~arcsec, 50~arcsec and 150~arcsec for the EMOS, EPN and FPM detectors, respectively. The physical scale of 10~kpc corresponds to 1.8~arcmin.}
    \label{img:regions}
\end{figure*}

Figure~\ref{img:regions} displays the X-ray images of NGC\,3982, with the top row presenting all \textit{XMM-Newton} camera images in the $0.3-10.0$\,keV band and the bottom row presenting the $3-78$\,keV images from each FPM camera onboard \textit{NuSTAR}. In all panels, the source\,$+$\,background extraction region is shown with a solid black line, whereas the background extraction regions are represented by a dashed line.


\section{X-ray spectral fitting}\label{sec:method}

The spectral analysis was performed for spectra binned using the \texttt{ftgrouppha} ftool\footnote{\href{https://heasarc.gsfc.nasa.gov/lheasoft/help/ftgrouppha.html}{https://heasarc.gsfc.nasa.gov/lheasoft/help/ftgrouppha.html}} such that each background spectrum would contain at least one count per bin, using the \texttt{grouptype\,=\,bmin} and \texttt{groupscale\,=\,1} options. The \textsc{Xspec} package \citep{Arnaud1996} v12.12.0g was then used for spectral fitting with the modified Cash statistic (aka W-stat; \citealt{Cash1979,Wachter1979}) invoked with the \texttt{cstat} command in \textsc{Xspec}. All datasets, including the three \textit{XMM-Newton} spectra and the two \textit{NuSTAR} spectra, were modelled simultaneously by tying all parameters apart from a pre-multiplying cross-calibration constant.

All spectral fits in this paper used the Bayesian X-ray Analysis (BXA, \citealp{Buchner2014}) software package for global parameter estimation, unless otherwise stated. BXA connects the \textsc{Xspec} Python interface (\textsc{PyXspec}) with nested sampling algorithms that iteratively update the global parameter distribution initially defined by the user with parameter priors. As such, the algorithm is very effective at avoiding local minima and traversing towards the global statistical minimum associated with the model being fit to the data. BXA is thus also capable of reproducing complex inter-parameter dependencies and complex parameter posterior shapes that describe the multi-dimensional parameter constraints attainable with a given model. In this work we use BXA v2.9 implemented with PyMultiNest \citep{Buchner2014}, the Python implementation of the nested sampling algorithm MultiNest \citep{Feroz2009}.


We performed the broadband X-ray spectral analysis in the energy range $0.3-10.0$~keV for the \textit{XMM-Newton} and $3-78$~keV for the \textit{NuSTAR} data by applying a variety of different spectral models, including both physically-motivated and the empirically-motivated phenomenological model \textsc{pexrav} \citep{Magdziarz1995}. In the case of \textsc{MYTorus} we performed the fitting starting at $0.6$~keV, as this is the minimum energy allowed by the model.

For all models, the fixed parameters throughout the work are as follows. The multiplicative constant (\texttt{C$_{\rm k}$}), corresponding to the instrument cross-calibration was initially left free to vary to check for significant spectral offsets. No significant deviations from unity were found, such that the cross-calibration was frozen to unity in all subsequent fits, in agreement with the values found by \citet{Madsen2017}. The redshift of NGC\,3982 was frozen to 0.00371 \citep{Martinsson2013}, while the Galactic column density in the direction of the source (\texttt{N$_{\rm H,\,gal}$}) obtained from the \texttt{nh} tool was frozen to $10^{20}\, \mathrm{cm^{-2}}$ \citep{HI4PI}. For all models we include an additional collisionally-ionised \texttt{apec} model component to reproduce the soft excess emission $\lesssim$3\,keV. The abundance of the \texttt{apec} component was set to Solar, while the temperature and normalisation were left free to vary limited to ranges [$0.01,5.0$]~keV and [$10^{-8},10^{-2}$]~keV\,$\rm{cm^{-2}\,s^{-1}}$. The Thomson-scattered `warm mirror' emission was modelled as a fraction of the intrinsic powerlaw continuum (\texttt{f$_{\rm scatt}$}), representing a fraction of the coronal emission that passes through an obscurer with lower column density than the primary obscurer. The fraction is expected to account for $\lesssim 10$\% of the intrinsic X-ray continuum in obscured AGN \citep{Gupta2021}. For all models where a \texttt{cutoffpl} component was used as the intrinsic continuum with variable high-energy cut-off, the value was frozen to 300\,keV in agreement with the value found by \citet{Balokovic2020}. For all free parameters discussed in \S\ref{sec:method} uniform or log-uniform priors were used, except for the photon index of the power-law component, where we defined a Gaussian prior with mean 2 and standard deviation 0.1, in agreement with previous results \citep{Ricci2017b}. 

\subsection{\textsc{pexrav}}
The phenomenological model \textsc{pexrav} \citep{Magdziarz1995} assumes an intrinsic exponentially cut-off power-law spectrum reflected from a neutral semi-infinite slab. Such a geometry is not physically relevant for the obscurer in AGN since it does not account for transmission through the material nor for reprocessing from the obscurer. However its inclusion provides an interesting comparison basis for the physical obscurer models we hitherto describe. In \textsc{Xspec} parlance, the model used was defined as:

\begin{center}
\texttt{C$_{\rm k}$\,$\times$\,N$_{\rm H,\,gal}$\,(apec\,$+$\,ztbabs\,$\times$\,cabs\,$\times$\,cutoffpl\,$+$ pexrav\,$+$\,f$_{\rm scatt}$\,$\times$\,cutoffpl\,$+$\,zgauss)}.
\end{center}

 The intrinsic obscurer\footnote{Here we refer to the intrinsic obscurer as the obscurer at the redshift of the source, to distinguish it from the Galactic obscurer. However, no assumption is made to decouple the host galaxy and circum-nuclear obscurers in our modelling.} was described by the model components \texttt{ztbabs\,$\times$\,cabs}, in which the column densities were tied and allowed to vary with a log-uniform prior between $10^{23}\,-\,10^{26}$\,$\rm cm^{-2}$. These two components reproduce the effects of photoelectric absorption and optically-thin Compton scattering, respectively. We note that these model components are well known to struggle to accurately reproduce the predicted spectrum at high column densities (e.g., \citealt{Yaqoob1997}), and such model limitations are exactly the scenarios we seek to compare with more appropriate physically-motivated models. The Fe\,K$\alpha$ line was modelled with a \texttt{zgauss} component with line energy and width frozen to 6.4\,keV and 1\,eV, respectively. The only free parameter for this component was hence the normalisation. The photon index and normalisation of the \texttt{pexrav} component were tied to the intrinsic continuum, while the cosine of the reflector inclination angle was left free to vary with a uniform prior in the range [$0.10,0.99$]. The relative reflection fraction of \texttt{pexrav} (\texttt{rel\textunderscore refl}) was limited to negative values (to reproduce the pure reflection spectrum) with a log-uniform prior in the range [$-100,-0.01$].

 \subsection{\textsc{Borus02}}
 \textsc{Borus02} is a physically-motivated obscuring torus model based on the Monte Carlo Radiative Transfer simulations of \citet{Balokovic2018}. The geometry of the reprocessing medium is a uniform-density sphere with two conical polar cutouts with variable half-opening angle to define the covering factor. The model was defined with the following expression:

\begin{center}
\texttt{C$_{\rm k}$\,$\times$\,N$_{\rm H,\,gal}$\,(apec\,$+$\,atable\{borus02\textunderscore v170323a\}\,$+$ zphabs\,$\times$\,cabs\,$\times$\,cutoffpl\,$+$\,f$_{\rm scatt}$\,$\times$\,cutoffpl).}
\end{center}

 We used the \textsc{borus02} model in both coupled and decoupled mode. For \texttt{borus02-coupled}, all three column densities corresponding to the \texttt{borus02}, \texttt{zphabs} and \texttt{cabs} components were tied together to vary log-uniformly in the range [$10^{22},10^{25.5}$]\,$\rm cm^{-2}$. The cosine of the inclination angle was left free to vary uniformly in the range [$18.2,84.3$]~degrees. For \texttt{borus02-decoupled} the line-of-sight column density of the \texttt{zphabs} and \texttt{cabs} components were tied together but allowed to vary independently from the torus column density of the \texttt{borus02} component. The inclination angle was frozen to $\approx84$~degrees (see \citealt{Lamassa2019} for more details of this setup). For both the decoupled and coupled model setup, the photon index and normalisation of the different components were tied together and left free to vary. The covering factor ($f_{\rm{C}}$) of the obscurer was also allowed to vary uniformly over the range [$0.1,0.99$].


 \subsection{\textsc{MYTorus}}
 \textsc{MYTorus} \citep{Murphy2009} describes an obscurer with a uniform density tube-like azimuthally symmetric torus corresponding to a classical \lq doughnut\rq\ geometry with a fixed opening angle of 60~degrees. The model consists of three components, each represented with a different table model. These tables are the zeroth-order continuum component \texttt{mytorus\textunderscore Ezero}, the Compton-scattered continuum component \texttt{mytorus\textunderscore scattered} and the fluorescence line emission \texttt{mytl}. We use the table models with termination energy of $E_\mathrm{T} = 300$\,keV and additionally include a Gaussian smoothing function for the fluorescent emission lines with the $\sigma_\mathrm{L}$ parameter via the \texttt{gsmooth} model component. The corresponding model expression was:

\begin{center}
\texttt{C$_{\rm k}$\,$\times$\,N$_{\rm H,\,gal}$\,(apec\,$+$\,f$_{\rm scatt}$\,$\times$\,zpowerlw\,$+$\,zpowerlw\,$\times$ etable\{mytorus\textunderscore Ezero\textunderscore v00\}\,$+$\,A$_{\rm S}$\,$\times$ atable\{mytorus\textunderscore scatteredH300\textunderscore v00\}\,$+$\,A$_{\rm L}$\,$\times$ gsmooth\,$\times$\,atable\{mytl\textunderscore V000010nEp000H300\textunderscore v00\}). }
\end{center}

 For \texttt{MYTorus-coupled}, we tied the photon index, normalisation, equatorial column density, inclination angle and intrinsic normalisation between all three table models. The equatorial column density was left free to vary with a log-uniform prior in the range [$10^{22},10^{25}$]\,$\rm cm^{-2}$, while the cosine of the inclination angle was left free to vary uniformly within the range corresponding to [$15,89$]~degrees. The width of the Fe K$\alpha$ line $\sigma_\mathrm{L}$ of the \texttt{gsmooth} component was frozen to $10^{-4}$~keV with energy index $\alpha$ fixed to one.

 \subsection{\textsc{UXCLUMPY}}
 The Unified X-ray \textsc{CLUMPY} (\textsc{UXCLUMPY}) torus model is based on the Monte Carlo Radiative Transfer simulation code \texttt{XARS}\footnote{Available at \href{https://github.com/JohannesBuchner/xars}{https://github.com/JohannesBuchner/xars}.} developed by \cite{Buchner2019}. The geometry of this physically-motivated table model is based on the clumpy torus model of \citet{Nenkova2008a} with $\mathcal{N}_\mathrm{tot} = 10^5$ spherical randomly distributed clouds. The obscurer column density is the highest in the equatorial plane whereas the number of clouds along the line-of-sight for edge-on system is $\mathcal{N}_0$. The number of clouds seen along the radial line-of-sight $\mathcal{N}$ is axis-symmetric and decreases with the inclination angle towards the poles. \textsc{UXCLUMPY} is the only model we employ that includes two distinct geometrical components. The torus dispersion of the main cloud population \texttt{TORsigma} effectively controls the torus scale height and its cosine was allowed to vary uniformly in the range corresponding to [$0,84$]~degrees. An additional inner ring of dense Compton-thick clouds is included in \textsc{UXCLUMPY} that was found to have a significant effect on the observed Compton hump profile by \citet{Buchner2019}. The \texttt{CTKcover} parameter describing the covering factor of that obscurer was allowed to vary uniformly in the range [$0.1,0.6$].

The model set-up is composed of three table models corresponding to the transmitted and cold reflected components with fluorescent line emission and the warm mirror component (the `omni' component). The line-of-sight column density $N_\mathrm{H,los}$ was allowed to vary log-uniformly in the range $[10^{23},10^{26}]\, \mathrm{cm^{-2}}$. The model expression was defined as:

\begin{center}
\texttt{C$_{\rm k}$\,$\times$\,N$_{\rm H,\,gal}$\,(apec\,$+$ atable\{uxclumpy-cutoff-transmit\}\,$+$ atable\{uxclumpy-cutoff-reflect\}\,$+$\,f$_{\rm scatt}$\,$\times$ atable\{uxclumpy-cutoff-omni\}).}
\end{center}
 
The parameters of all table components were tied together and the inclination angle was allowed to vary in the range [$0,90$]~degrees.

 \subsection{\textsc{RXTorus}}
 \textsc{RXTorus} is a physically-motivated model by \cite{Paltani2017}, generated with the \texttt{RefleX} platform\footnote{Available at \href{https://www.astro.unige.ch/reflex}{https://www.astro.unige.ch/reflex}}, a Monte Carlo code designed for tracking the propagation of individual X-ray photons through distributions of gas and dust. \textsc{RXTorus} is the first application of \texttt{RefleX}, adapting the same source and absorber geometries as \textsc{MYTorus} whilst including the covering factor as a free parameter. The model consists of three table model components; the \texttt{RXTorus\textunderscore cont\textunderscore M} is an exponential multiplicative tabular model for the continuum absorption component where \texttt{M} denotes the metallicity (we assumed Solar with \texttt{M}~=~1). The reprocessed emission includes the Compton scattered and fluorescent line emission and is given by the \texttt{RXTorus\textunderscore rprc\textunderscore M\textunderscore CCC} component where \texttt{CCC} describes the high energy cut-off. The model was defined using the following expression:
 
\begin{center}

\texttt{C$_{\rm k}$\,$\times$\,N$_{\rm H,\,gal}$\,(apec\,$+$\,f$_{\rm scatt}$\,$\times$\,cutoffpl\,$+$ A$_{\rm C}$\,$\times$\,cutoffpl\,$\times$\,etable\{RXTorus\textunderscore cont\textunderscore 1\}\,$+$ A$_{\rm S}$\,$\times$\,atable\{RXTorus\textunderscore rprc\textunderscore 1\textunderscore 200\}).}
\end{center}

 In \textsc{RXTorus}, the line-of-sight column density is defined by the following expression \citep{Paltani2017}:

 \begin{center}
     \begin{equation}
         N_\mathrm{H,los}(\theta_\mathrm{i}) = N_\mathrm{H,eq} \left( 1 - \frac{R^2}{r^2} \cos^2{\theta_\mathrm{i}} \right)^{1/2}, \hspace{1ex} \cos{\theta_\mathrm{i}} < \frac{r}{R},
     \end{equation}
 \end{center}

 where $\theta_\mathrm{i}$ represents the inclination angle and 
 for $\cos{\theta_\mathrm{i}} > r/R$ the line-of-sight column density is equal to zero. In our analysis, the equatorial column density $N_\mathrm{H,eq}$ was left free to vary log-uniformly in the range [$10^{23},10^{25}$]\,$\rm cm^{-2}$. The ratio between the minor and major torus radii, $r/R$, represents the covering factor of the torus and was left as a free parameter to vary uniformly in the range [$0.1,0.8$], while the inclination angle was allowed to vary in the range [$20,89$]~degrees. All parameters were tied between individual table models and the intrinsic continuum assumed an exponential cut-off energy of 200~keV.

\section{Results}\label{sec:results}

\begin{figure*}[t]
    \includegraphics[width=\textwidth]{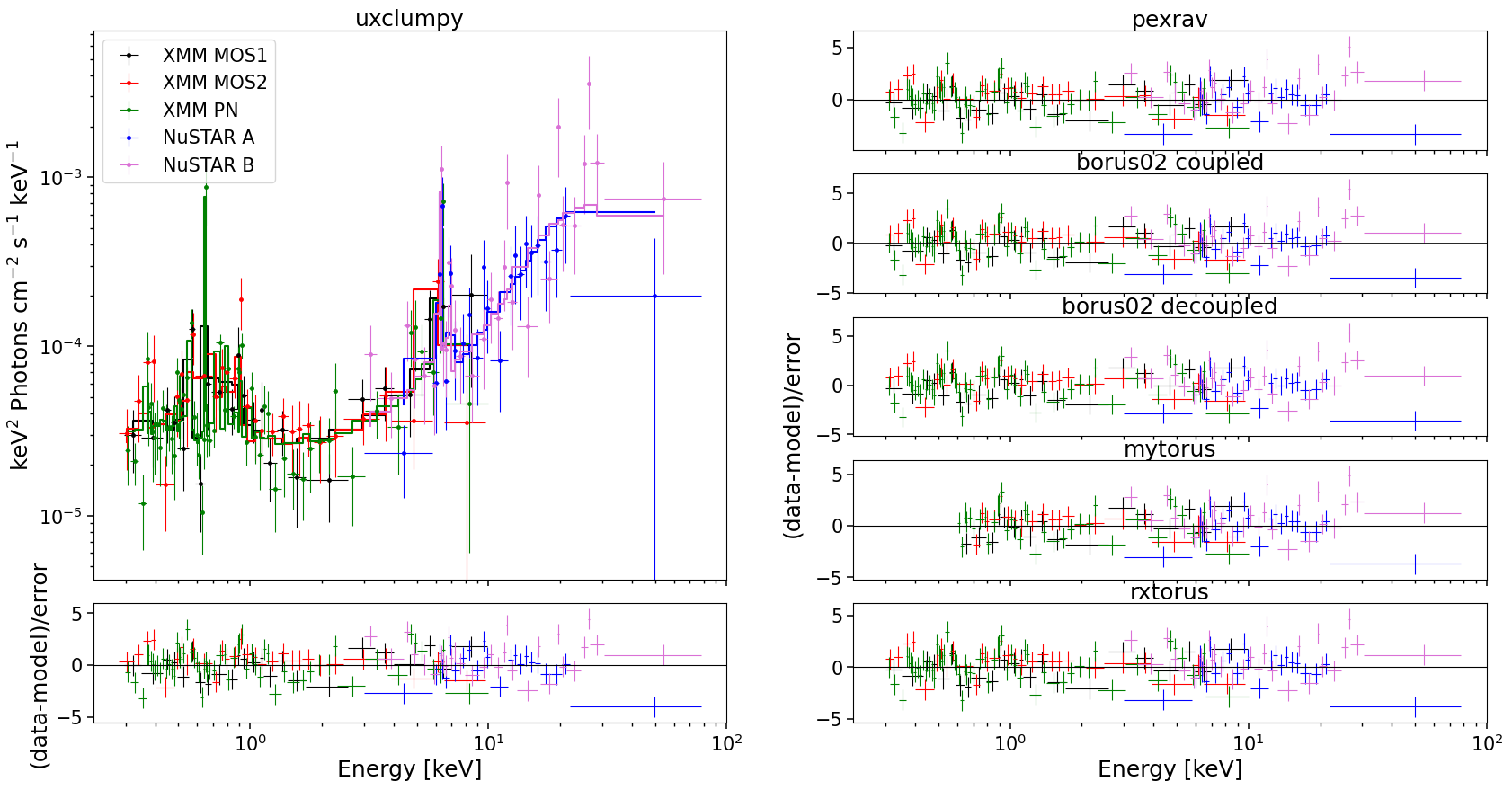}
    \caption{The \textit{XMM-Newton} and \textit{NuSTAR} spectra modelled and unfolded with \textsc{UXCLUMPY} model (\textit{top left}) with corresponding residuals (\textit{bottom left}) and residuals for the rest of the tested spectral models (\textit{right}) described in \S\ref{sec:method}.}
    \label{img:xspec bestfit}
\end{figure*}

Initially we fit the broadband X-ray spectra with the default minimization algorithm available in \textsc{Xspec} (the Levenberg-Marquardt algorithm; \citealt{Levenberg1944,Marquardt1963}), which converges towards a minimum using local information from the surrounding parameter space starting at an initial parameter guess. The fit with the \textsc{UXCLUMPY} model obtained within \textsc{Xspec} is shown in Figure~\ref{img:xspec bestfit}, together with the corresponding residuals (\textit{bottom left} panel) and the residuals obtained for all the different models tested here (\textit{right} panels). A visual comparison of the residuals shows how indistinguishable the model fits are for all data sets. Similarly the fit statistics ($C$/d.o.f.) listed in Table \ref{tab:results bxa} shows that the fits are comparable among all models. However, the derived posteriors for some of the parameters varied dramatically among different models. For instance, we found large discrepancies between the derived scattering fraction, ranging from 0.01\% found by \texttt{borus02-decoupled} up to 10\% estimated by \textsc{pexrav}. The rest of the models predict scattering fraction in range 0.1-0.5\%. The covering factor and the inclination angle were also found to differ dramatically, predicting the opening angle in range 60 degrees up to 84 degrees and inclination from 21 degrees up to almost edge-on system by different models. On the other hand, the column density for most of the models was found to be pegged to the upper limits defined by the specific model, suggesting the Compton-thick nature of NGC\,3982. The fundamental reason for such differences is the ability of models to reproduce the same spectral shapes with unique parameter combinations.


We next performed the analysis using BXA for all described models. Results from the global parameter estimation, showing the median and uncertainties at 68\% confidence level of the posterior estimates, are shown in Table~\ref{tab:results bxa}, together with the Bayesian global evidence ln $Z$. All models describe the observed spectra similarly well, though we note the Bayesian evidence for \textsc{MYTorus} is not listed since we fit this model to data with energy $E$\,$>$\,0.6 keV. Corner plots displaying all posterior probability distributions obtained by BXA are shown in Appendix \ref{appendix:corners}.

There is broad agreement between the line-of-sight column densities derived by the different models, all of them resulting in $N_{\rm H}>1.5\times10^{24}\,\rm{cm^{-2}}$ to $>99\%$ confidence, thus confirming the Compton-thick nature of the circum-nuclear obscurer in NGC\,3982. Thefrom equatorial column density in both \textsc{MYTorus} and \textsc{RXTorus} and the global column density in \texttt{borus02-decoupled} were pegged to the maximum value allowed for this parameter.

While the covering factor of the coupled \textsc{MYTorus} model is fixed to a value of 0.5 (i.e. a half-opening angle of 60 degrees), this parameter was free in the other models. For \textsc{RXTorus} we found a covering factor of $0.47\pm0.12$, corresponding to a half-opening angle of $62\pm8$~degrees. For the \textsc{UXCLUMPY} model we found \texttt{TORsigma}\,=\,$61^{+13}_{-20}$~degrees and the Compton-thick covering factor of the inner ring, \texttt{CTKcover}\,=\,$0.2\pm0.1$. Owing to the probabilistic generation of clouds from a distribution \citep{Buchner2019}, the covering factor in \textsc{UXCLUMPY} is difficult to analytically define. We estimated the covering factor using a pre-computed grid of \texttt{TORsigma}, \texttt{CTKcover} and covering factor for sightlines with log\,$N_{\rm H}$\,$>$\,24 \citep{Boorman2023}. We used a grid interpolator on a random sample of \textsc{UXCLUMPY} posterior rows to produce a posterior on covering factor given by $f_{\rm C}\approx0.5\pm0.1$, which is fully consistent with the values found by the other models (see Figure~\ref{img:degs}).

We find that for \texttt{borus02-coupled} the preferred inclination angle is smaller than the half-opening angle of the torus at 3$\sigma$ confidence level, which might suggest an unobscured source. However, the line-of-sight component of this model is not dependent on the inclination angle or on the very high global column density ($\approx10^{25}\,\rm{cm^{-2}}$). Inclination angle smaller than the half-opening angle could suggest a scenario in which the spectrum is dominated by reflection from the back wall of the torus, in favour with the reflection-dominated broadband spectrum. On the other hand, for \texttt{borus02-decoupled} we obtain an upper limit of the covering factor ($f_{\rm C}\lesssim 0.17$). This indicates that the fit favours a dense low covering obscurer possibly resembling a disc-like structure. It should be noted, however, that this might also be caused by our assumption that the inclination angle corresponds to an edge-on system. Additionally, further uncertainties may arise from an error in the Green functions associated with \textsc{borus02} as reported by \cite{VanderMeulen2023}, though we note the line-of-sight column density contours from \textsc{borus02} are in good agreement with with the other models.

\subsection{Intrinsic AGN luminosity \& Eddington ratio}

The posterior values of the photon index and power-law normalisation were used to determine the intrinsic luminosity using the Hubble distance corrected to the Reference Frame defined by the 3K CMB, $18.91$~Mpc\footnote{\href{https://ned.ipac.caltech.edu}{https://ned.ipac.caltech.edu}}. The observed median $2-10$~keV luminosity was found to be $\approx(4.6-5.9)\times10^{39}\,\rm{erg\,s^{-1}}$ across different models. We further estimated the intrinsic X-ray luminosity in two different energy bands, $2-10$~keV and $8-24$~keV, respectively, as listed in Table~\ref{tab:results bxa}. The intrinsic luminosity of \texttt{borus02-coupled} was found to have the lowest median value $L_{2-10\,\rm keV}\approx8\times10^{40}\,\rm{erg\,s^{-1}}$, which was up to two orders of magnitude lower than the highest one found by \texttt{borus02-decoupled}, with median $L_{2-10\,\rm keV}\approx1\times10^{42}\,\rm{erg\,s^{-1}}$. However, for the majority of the models, the median $2-10$~keV intrinsic luminosity was found to be in the range $\approx(1.9-3.7)\times10^{41}\,\rm{erg\,s^{-1}}$.

The Eddington ratio ($\lambda_{\rm Edd}$) is an important parameter as it traces the growth rate of SMBHs. It is given as a fraction of the bolometric luminosity and the Eddington luminosity, $\lambda_{\rm Edd}=L_{\rm bol}/L_{\rm Edd}\sim L_{\rm bol}/M_{\rm BH}$, where $M_{\rm BH}$ is the mass of the SMBH. We calculated the bolometric luminosity of NGC\,3982 using the $2-10$~keV band luminosity adopting the X-ray bolometric correction for Compton-thick AGN from \cite{Brightman2017}, $\rm{log}\,\kappa_{2-10}=1.44\pm0.12$. The Eddington luminosity was derived via the black hole mass given in \citeauthor{Kammoun2020} (\citeyear{Kammoun2020}; $\mathrm{log}\,M_{\mathrm{BH}}=6.89~M_\odot$), which was calculated from the M-sigma relation from \cite{KormendyHo2013}. We included an additional 0.5~dex uncertainty for all subsequent Eddington ratio estimations. The median Eddington ratios we find encompass the range $\approx0.2-1.1$\,\% for most of the models, however for \textsc{pexrav} and \texttt{borus02-decoupled} the estimated Eddington ratio posteriors covered substantially larger ranges, as high as $6$\% at the 68\% confidence level. Both bolometric luminosities and Eddington ratios for all models are listed in Table~\ref{tab:results bxa}.

Regarding the soft X-ray band, all models found very similar values for the temperature and normalisation of the \texttt{apec} component. The temperature was found to be $\approx\,0.28$~keV for the majority of the models with a median normalisation in the range $\approx(4.2-4.5)\times10^{-5}\,\mathrm{keV\,cm^{-2}\,s^{-1}}$. Only \texttt{MYTorus-coupled} found slightly higher temperatures of $0.33-0.43$~keV and median normalisation $\approx2.9\times10^{-5}\,\mathrm{keV\,cm^{-2}\,s^{-1}}$, though it is important to stress that \textsc{MYTorus} is limited in the soft X-ray band to energies $>$0.6~keV. We further estimated the intrinsic soft X-ray luminosity in the $0.5-2.0$~keV band from the \texttt{apec} component to derive the host-galaxy star-formation rate, using equation~2 in \citet{Mineo2012b}. We note that \citet{Mineo2012b} calculated luminosities with \texttt{mekal}, though the differences with our \texttt{apec}-based measurements should be minimal given the statistical uncertainty of the data being fit. The soft X-ray luminosity derived from the \texttt{apec} model component was found to be $\approx(3.2\pm0.3)\times10^{39}\,\rm erg\,s^{-1}$, while the corresponding star-formation rates were found to be $\approx6\pm2\,\rm{M_\odot\,yr^{-1}}$ for the majority of the models. The \textsc{MYTorus} model predicts a slightly smaller luminosity of $\approx(2.6\pm0.3)\times10^{39}\,\rm erg\,s^{-1}$, corresponding to a predicted star-formation rate of $\approx5\pm2\,\rm{M_\odot\,yr^{-1}}$. 
These results are consistent within the uncertainties with findings of \cite{Lianou2019}, who estimated a global star-formation rate of $\approx2.9\pm2.7~\rm{M_\odot\,yr^{-1}}$, using the \textit{WISE} 22$\mu$m band corrected for emission of evolved stars.

\begin{table*}[ht]
\caption{Posterior parameters constraints with all uncertainties corresponding to 68\% confidence level.}
\label{tab:results bxa}
\begin{center}
\begin{threeparttable}
\makebox[\paperwidth]%
{
\renewcommand{\arraystretch}{1.6}
\hspace{-6.1cm}%
\begin{tabular}{llccccccc}

\hline

Component & Parameter & Unit & \textsc{pexrav} & \textsc{borus02} c. & \textsc{borus02} d. & \textsc{MYTorus} & \textsc{UXCLUMPY} & \textsc{RXTorus}  \rule{0pt}{2.6ex} \\ [1ex]
\hline \hline

\textsc{apec} & $\log L^{\rm{int}}_\mathrm{.5-2.}$ & $\rm{erg\,s^{-1}}$ & $39.52^{+0.04}_{-0.04}$ & $39.50^{+0.04}_{-0.04}$ & $39.50^{+0.04}_{-0.05}$ & $39.41^{+0.05}_{-0.05}$ & $39.50^{+0.04}_{-0.04}$ & $39.50^{+0.04}_{-0.04}$  \rule{0pt}{3.2ex} \\ [1ex]
& $\rm{SFR}_{\rm{X-ray}}$ & $\rm{M_\odot\,yr^{-1}}$ & $6.4^{+2.2}_{-2.3}$ &  $6.1^{+2.2}_{-2.2}$ & $6.0^{+2.3}_{-2.1}$ & $5.0^{+2.2}_{-2.2}$ & $6.1^{+2.2}_{-2.2}$ &  $6.1^{+2.1}_{-2.2}$ \rule{0pt}{3.2ex} \\ [1ex]

\hline

Torus properties & $\log N_\mathrm{H,los}$ & $\rm{cm^{-2}}$ & $25.23^{+0.51}_{-0.53}$ & $25.06^{+0.29}_{-0.45}$ & $24.79^{+0.14}_{-0.12}$ & $24.74^{+0.08}_{-0.09}$ & $25.35^{+0.43}_{-0.45}$ & $24.75^{+0.09}_{-0.11}$ \rule{0pt}{3.2ex} \\ [1ex]
& $\log N_\mathrm{H,eq}$ & $\rm{cm^{-2}}$ & ... & ... & $24.96^{+u}_{-0.68}$ & $24.97^{+u}_{-0.05}$ & ... & $24.95^{+u}_{-0.06}$ \rule{0pt}{3.2ex} \\ [1ex]
& $f_\mathrm{C}$ & ... & ... & $0.51^{+0.17}_{-0.18}$ & $0.14^{+0.03}_{-u}$ & 0.5\tnote{a} & $0.52^{+0.08}_{-0.09}$\tnote{$\star$} & $0.47^{+0.12}_{-0.12}$ \rule{0pt}{3.2ex} \\ [1ex]
& $\theta_{\rm{half-opening}}$ & deg & ... & $60^{+11}_{-13}$ & $82^{+u}_{-2}$ & 60\tnote{a} & $59^{+6}_{-6}$\tnote{$\star$} & $62^{+8}_{-8}$ \rule{0pt}{3.2ex} \\ [1ex]
& $\theta_{\rm inclination}$ & deg. & $44^{+18}_{-19}$ & $41^{+14}_{-13}$ & 84.3\tnote{a} & $67^{+3}_{-2}$ & $60^{+21}_{-28}$ & $70^{+7}_{-9}$ \rule{0pt}{3.2ex} \\ [1ex]

\hline

Scattering fraction & $f_\mathrm{scatt}$ & \% & $0.88^{+3.83}_{-0.76}$ & $3.30^{+1.31}_{-1.13}$ & $0.22^{+0.15}_{-0.09}$ & $1.56^{+0.32}_{-0.32}$ & $7.93^{+6.20}_{-3.54}$ & $0.66^{+0.29}_{-0.21}$ \rule{0pt}{3.2ex} \\ [1ex]

\hline

Power-law & $\Gamma$ & ... & $1.96^{+0.09}_{-0.08}$ & $1.95^{ +0.08}_{-0.07}$ & $1.98^{+0.09}_{-0.08}$ &$1.89^{+0.09}_{-0.09}$  & $2.03^{+0.07}_{-0.07}$ & $2.00^{+0.01}_{-0.01}$ \rule{0pt}{3.2ex} \\ [1ex]
& log\,$K$ & keV\,$\rm{cm^{-2}\,s^{-1}}$ & $-2.62^{+0.87}_{-0.73}$ & $-3.15^{+0.17}_{-0.14}$ & $-1.98^{+0.26}_{-0.22}$ & $-2.82^{+0.08}_{-0.07}$ & $-2.61^{+0.15}_{-0.14}$ & $-2.47^{+0.16}_{-0.15}$ \rule{0pt}{3.2ex} \\ [1ex]

\hline

AGN properties & $\log F^{\rm{obs}}_\mathrm{2-10}$ & $\rm{erg\,cm^{-2}\,s^{-1}}$ & $-12.71^{+0.02}_{-0.02}$ & $-12.71^{+0.02}_{-0.02}$ & $-12.71^{+0.02}_{-0.02}$ & $-12.82^{+0.06}_{-0.05}$ & $-12.73^{+0.02}_{-0.02}$ & $-12.74^{+0.02}_{-0.02}$   \rule{0pt}{3.2ex} \\ [1ex]
& $\log F^{\rm{int}}_\mathrm{2-10}$ & $\rm{erg\,cm^{-2}\,s^{-1}}$ & $-11.17^{+0.86}_{-0.73}$ & $-11.72^{+0.16}_{-0.11}$ & $-10.57^{+0.25}_{-0.19}$ & $-11.35^{+0.10}_{-0.08}$ & $-11.23^{+0.12}_{-0.11}$ & $-11.06^{+0.16}_{-0.15}$  \rule{0pt}{3.2ex} \\ [1ex]

& $\log L^{\rm{obs}}_\mathrm{2-10}$ & $\rm{erg\,s^{-1}}$ & $39.77^{+0.02}_{-0.02}$ & $39.77^{+0.02}_{-0.02}$ & $39.77^{+0.03}_{-0.03}$ & $39.66^{+0.05}_{-0.05}$ & $39.75^{+0.02}_{-0.02}$ & $39.74^{+0.02}_{-0.02}$   \rule{0pt}{3.2ex} \\ [1ex]

& $\log L^{\rm{int}}_\mathrm{2-10}$ & $\rm{erg\,s^{-1}}$ & $41.46^{+0.86}_{-0.73}$ & $40.91^{+0.16}_{-0.11}$ & $42.07^{+0.25}_{-0.19}$ & $41.29^{+0.10}_{-0.08}$ & $41.40^{+0.12}_{-0.11}$ & $41.57^{+0.16}_{-0.15}$  \rule{0pt}{3.2ex} \\ [1ex]
& $\log L^{\rm{int}}_\mathrm{8-24}$ & $\rm{erg\,s^{-1}}$ & $41.30^{+0.85}_{-0.73}$ & $40.77^{+0.17}_{-0.11}$ & $41.91^{+0.23}_{-0.18}$ & $41.17^{+0.13}_{-0.11}$ & $41.22^{+0.12}_{-0.10}$ & $41.40^{+0.17}_{-0.15}$  \rule{0pt}{3.2ex} \\ [1ex]
& $\mathrm{log} \ L_{\mathrm{bol}}$ & $\rm{erg\,s^{-1}}$ & $42.89^{+0.87}_{-0.72}$ & $42.36^{+0.20}_{-0.17}$ & $43.51^{+0.26}_{-0.26}$ & $42.73^{+0.15}_{-0.15}$ & $42.85^{+0.16}_{-0.16}$ & $43.02^{+0.20}_{-0.18}$ \rule{0pt}{3.2ex} \\ [1ex]

& $\lambda_\mathrm{Edd}$ & \% & $0.80^{+5.04}_{-0.64}$ & $0.23^{+0.14}_{-0.08}$ & $3.32^{+2.71}_{-1.48}$ & $0.55^{+0.23}_{-0.16}$ & $0.72^{+0.33}_{-0.23}$ & $1.06^{+0.61}_{-0.37}$ \rule{0pt}{3.2ex} \\ [1ex]

\hline

Fit statistic & 
$C$/d.o.f. & ... &  \nicefrac{2354.9}{2877} & \nicefrac{2351.0}{2877} & \nicefrac{2351.7}{2877} & \nicefrac{2277.8}{2808} & \nicefrac{2352.1}{2876} & \nicefrac{2352.3}{2877} \rule{0pt}{3.2ex} \\ [1ex]
& $\mathrm{ln}\,Z$ & ... & $-1190.4$ & $-1186.3$ & $-1190.4$ & ... & $-1185.7$ & $-1188.3$ \rule{0pt}{3.2ex} \\ [1ex]

\hline \hline

\end{tabular}%
}
    \begin{tablenotes}
        \item[a] Fixed for the modelling.
        \item[\textit{u}] Uncertainty unconstrained as the fit is pegged to the hard limit.
        \item[$\star$] Estimated by linear grid interpolation using \texttt{TORsigma} and \texttt{CTKcover} for $\mathrm{log}\,N_{\rm H} > 24$.
    \end{tablenotes}
\end{threeparttable}
\end{center}
\end{table*}

\section{Discussion and Future prospects}\label{sec:discussion}

Given the majority of AGN in the Universe are obscured (\citealt{Ueda2014, Buchner2014, Brandt2015, Ricci2017b}), to understand the bulk of AGN we need to fully comprehend the intrinsic properties of those systems that are deeply buried in thick layers of dust and gas. By exploring the posterior parameter space of a Compton-thick low-luminosity AGN with different physically-motivated models, we can quantify how well different parameters of interest can be estimated reliably in low flux targets.

\begin{figure*}[t]
    \begin{center}
    \vspace{-2em}
    \includegraphics[width=0.4\textwidth]{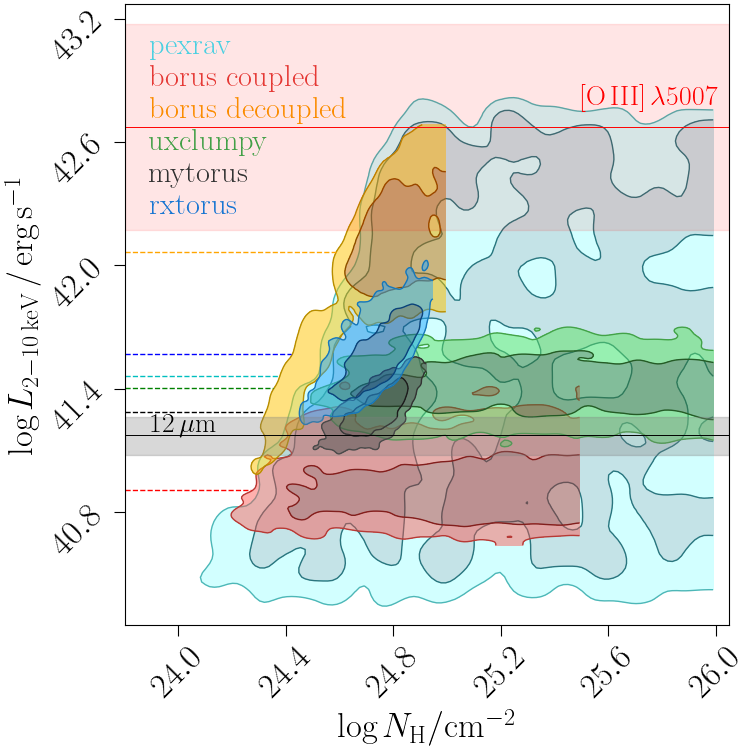}\hspace{2em}
    \includegraphics[width=0.4\textwidth]{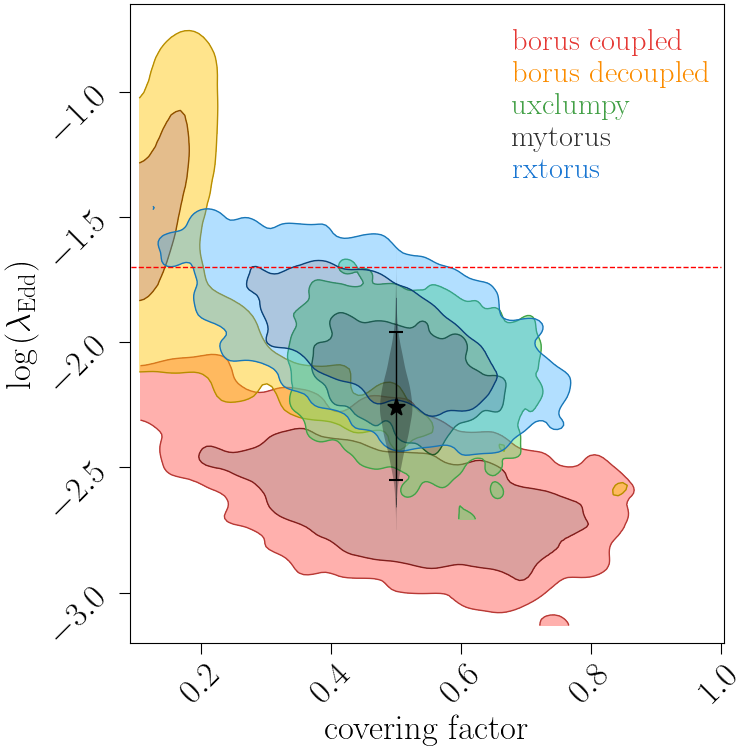}
    \includegraphics[width=0.4\textwidth]{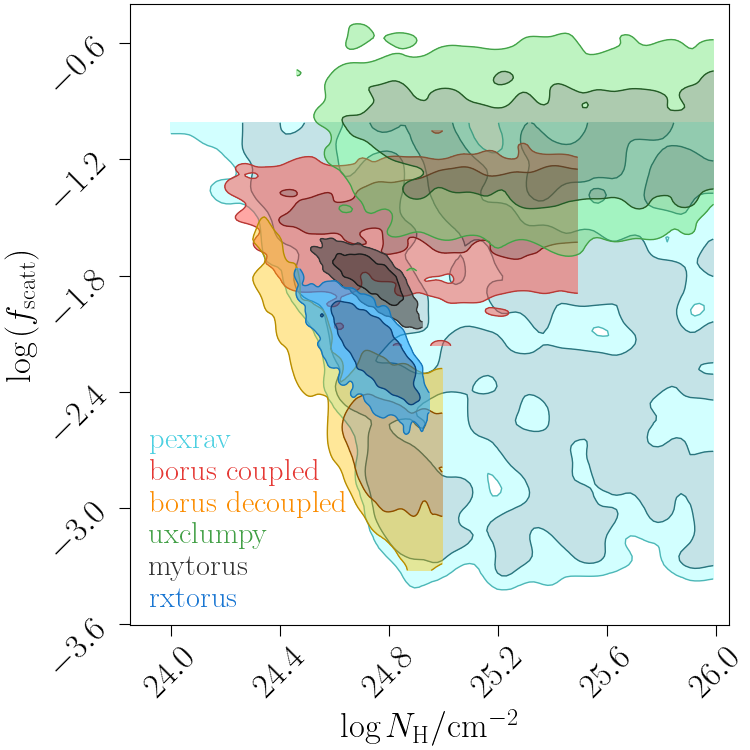}\hspace{2em}
    \includegraphics[width=0.4\textwidth]{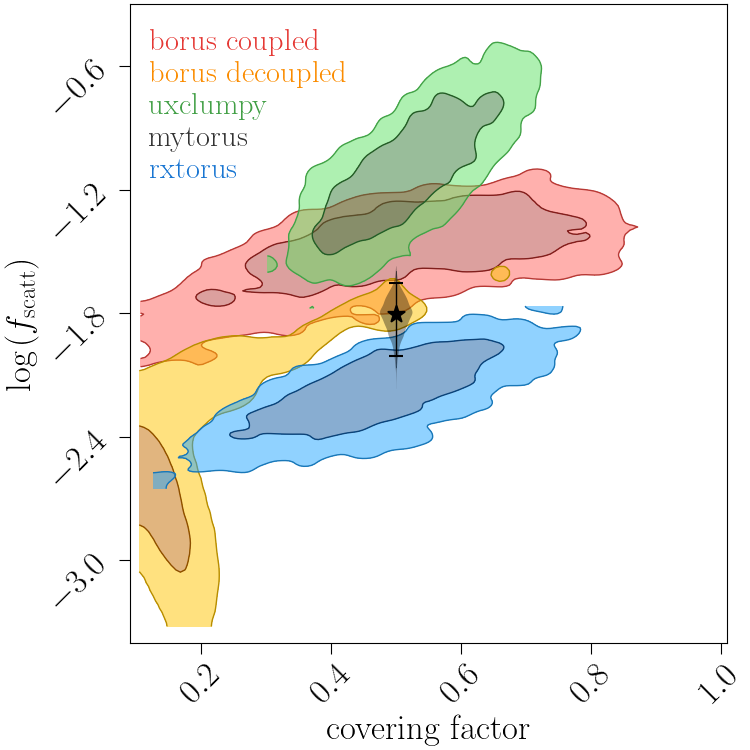}
    \includegraphics[width=0.4\textwidth]{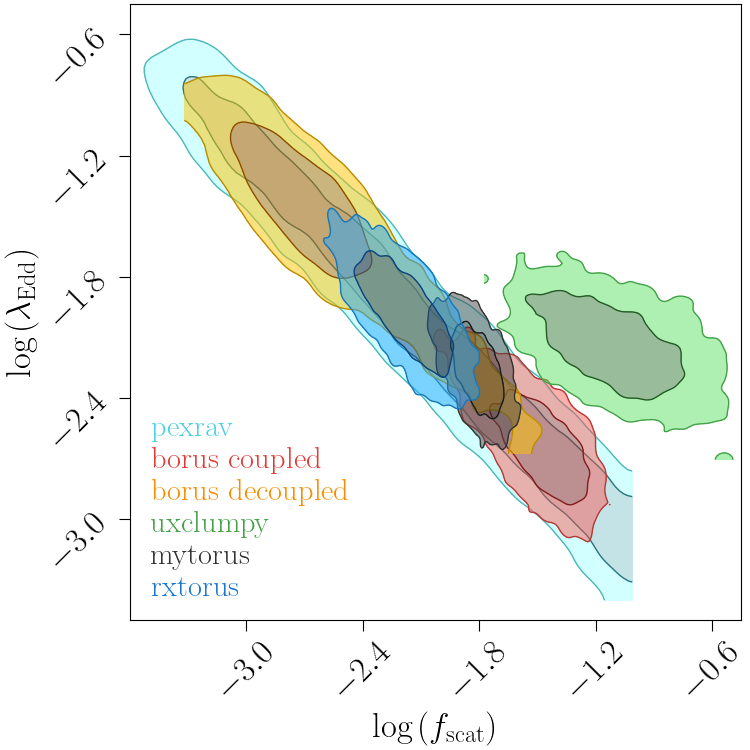}\hspace{2em}
    \includegraphics[width=0.4\textwidth]{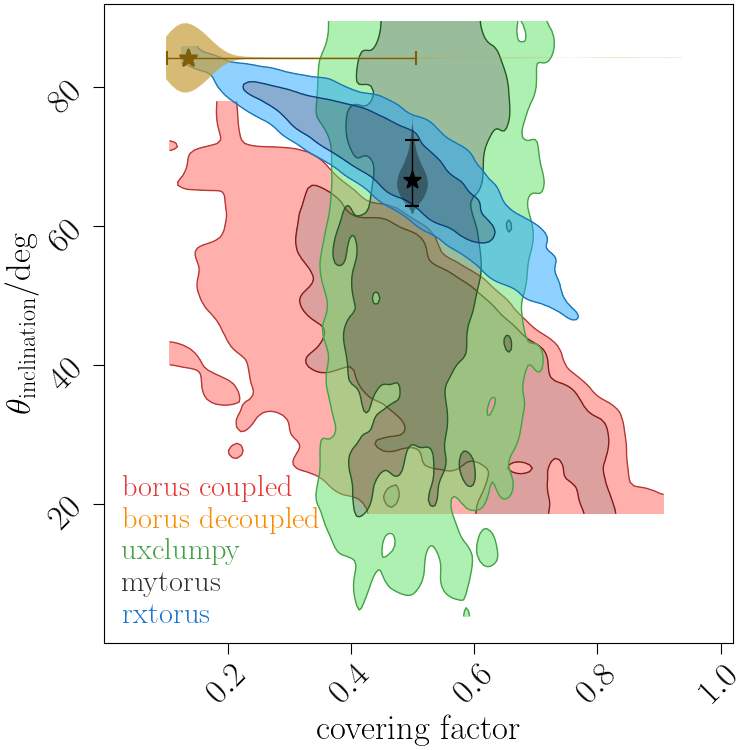}
    \caption{Contour plots for different posterior combinations. \textit{Top left panel}: The black and red solid lines represent 50th quantile of the intrinsic luminosity derived from Equations \ref{eq:12um to X-ray lumin} and \ref{eq:X-ray to OIII lumin}, respectively, with the shaded regions corresponding to 16th and 84th quantiles of the distributions. The dotted line shows the median luminosity for different models. \textit{Right panels}: the \textsc{MYTorus} model has covering factor fixed, the corresponding Eddington ratio distribution is shown with the violin plot while the star indicates the median, the errorbars indicate 2$\sigma$ uncertainties. Similarly, in \textit{bottom right panel} the inclination angle for \texttt{borus-decoupled} was fixed, so the corresponding covering factor distribution is shown with a violin plot and a 2$\sigma$ errorbar. \textit{Top right panel}: the red dashed line represents the effective Eddington limit for a dusty gas \citep{Ricci2017c}. Different contours show the 1$\sigma$ and 2$\sigma$ uncertainties.}
    \label{img:degs}
    \end{center}
\end{figure*}

The model posterior distributions obtained via global parameter estimation with BXA were used to visualise the corresponding parameter spaces of each model in Figure~\ref{img:degs}. Some models are seen to cover very large ranges, e.g. \textsc{pexrav} which covers a range of $\sim$2~dex for both intrinsic luminosity and line-of-sight column density. In some cases different geometries predict very different overall values for a given parameter, e.g. for \texttt{borus-coupled} 72\% of the posterior intrinsic luminosity points are below $10^{41} \, \rm ergs\,s^{-1}$, however, for \textsc{pexrav} it is only 29\% and other models predict the intrinsic luminosity higher than $10^{41} \, \rm ergs\,s^{-1}$ with $>$99\% probability. Similarly, regarding the scattering fraction, \texttt{borus-decoupled} and \textsc{RXTorus} predict the fraction of the scattered light smaller than 1\% with $>$86\% probability. On the other hand, \texttt{borus-decoupled}, \textsc{MYTorus} and \textsc{UXCLUMPY} predict scattering fraction higher than 1\% with $>$97\% probability, while for \textsc{pexrav} this value is close to the median  of the distribution. 

\subsection{Mid-infrared and Optical emission}
The intrinsic AGN X-ray luminosity can also be estimated by looking at emission in other `isotropic' wavelengths. For example, the primary radiation of an AGN can be re-emitted in the mid-infrared (MIR) waveband after being reprocessed by hot dust. The monochromatic $12\,\mu \rm m$ MIR luminosity is therefore found to be tightly correlated with the $2-10$~keV X-ray luminosity (\citealt{Elvis1978, Glass1982, Krabbe2001, Lutz2004, RamosAlmeida2007, Gandhi2009,Asmus2015}). To estimate the intrinsic X-ray luminosity from the nuclear MIR dust emission we adopted the following relation from \cite{Asmus2015}:

\begin{equation}
\log L_\mathrm{12\,\mu \rm m}^{43} = (0.97 \pm 0.03) \times \log L_\mathrm{2-10\,keV}^{43} + (0.33 \pm 0.04).
\label{eq:12um to X-ray lumin}
\end{equation}

\noindent
Here $L_\mathrm{12\,\mu \rm m}^{43}$ and $L_\mathrm{2-10\,keV}^{43}$ represent the luminosity in units of $10^{43} \, \rm{erg \, s^{-1}}$. Similarly, the correlation between the X-ray $2-10$~keV and optical [O\,III]\,$\lambda$5007 emission is also well explored (\citealt{Ward1988, Panessa2006, GonzalezMartin2009,Berney2015}), connecting the emission of the central regions to the extended partially ionized narrow-line region. We adopted the relation between optical [O\,III]\,$\lambda$5007 and $2-10$~keV X-ray luminosity from \cite{Berney2015}:

\begin{equation}
\log L_\mathrm{[O\,III]} = (1.23 \pm 0.05) \times \log L_\mathrm{2-10\,keV} + (-12 \pm 2).
\label{eq:X-ray to OIII lumin}
\end{equation}

\noindent
However, we note the [O\,III] can be strongly affected by variability as it traces the power of the central engine in the past, as well as host galaxy contamination that can lead to substantial scatter (e.g., \citealt{Ueda2003}). 
We therefore obtain a complementary estimate of the X-ray luminosity of NGC\,3982 by using both optical [O\,III]\,$\lambda5007$ and MIR $12\,\mu \rm m$ observations\footnote{We note that the [O\,III] and $12\,\mu \rm m$ luminosities were estimated using different luminosity distances. The maximum difference in derived $2-10$~keV luminosity arising from the discrepant distances does not exceed 0.1 dex.}. The nuclear MIR luminosity of NGC\,3982, $\log L_\mathrm{12\,\mu \rm m}=41.56\pm0.06$, was adopted from \cite{Asmus2015} and the [O\,III]\,$\lambda$5007 luminosity of NGC\,3982 $\log L_\mathrm{[O\,III]} = 40.50$, corrected for Galactic absorption and narrow line region extinction was adopted from \cite{Panessa2006}. The estimated $2-10$~keV luminosities using Equations~\ref{eq:12um to X-ray lumin}~and~\ref{eq:X-ray to OIII lumin} are plotted as horizontal lines and associated 68\% interquartile shaded range in the top left panel of Figure~\ref{img:degs}.

The [O\,III]\,$\lambda5007$ emission predicts $\log L_\mathrm{2-10\,keV} = 42.7^{+2.4}_{-2.3}\,\mathrm{erg\,s^{-1}}$. As notable from the top left panel of Figure~\ref{img:degs}, we find significantly lower intrinsic X-ray luminosity for most of the models, despite the large uncertainties arising from the adopted correlation{\footnote{We note that with additional scatter the [O\,III] luminosity may result consistent with results from our analysis.}. This suggests that the AGN might have been more active in the past, hence the higher X-ray luminosity inferred from the extended [O\,III] emission. This finding is in agreement with the discovery of \cite{Esparza-Arredondo2020}, who identified NGC\,3982 as a fading AGN candidate. On the other hand, from the MIR-versus-X-ray luminosity correlation we found $\log L_\mathrm{2-10\,keV}=41.2\pm0.1\,\mathrm{erg\,s^{-1}}$. This value is reproduced by all the tested models within 2$\sigma$ uncertainties, however we note some (e.g. \texttt{borus-decoupled} and \textsc{pexrav}) cover a very large range of predicted X-ray luminosity. In estimating the X-ray luminosity, the nuclear $12\,\mu \rm m$ emission is more precise than the [O\,III] line emission since the MIR-versus-X-ray relation has lower overall observed scatter. In addition, the MIR emission is likely being emitted from closer regions to the central engine than the [O\,III] emission arising on larger Narrow Line Region scales, and so may be less affected by host-galaxy contamination. Future multi-wavelength physically-motivated obscurer models such as those attainable with RefleX and SKIRT (e.g., \citealt{Andonie2022,Ricci2023,VanderMeulen2023}) would hence be useful in defining multi-wavelength prior constraints on the intrinsic AGN power derived in X-ray spectral fitting.

\subsection{Unique model dependencies}
As NGC\,3982 is a low-flux source, the uncertainties of posterior parameters are large, as illustrated in Figure~\ref{img:degs}. Even though the derived distributions cover similar ranges of parameter values (e.g. the lower limit of the column density is in agreement for all models), the shapes of the distributions in all two-dimensional posterior parameter spaces considered in Figure~\ref{img:degs} differ significantly. The distributions in the top left panel of Figure~\ref{img:degs} illustrate how for higher column density a larger intrinsic luminosity is needed. This is expected as for more luminous sources a higher line-of-sight column would be required to reproduce the same observed flux. The right panels of Figure~\ref{img:degs} show the parameter posterior dependencies between covering factor and intrinsic power propagated into Eddington ratio, scattering fraction and inclination angle, which are found to be somewhat degenerate across all models. Overall the covering factor is only poorly constrained with large uncertainties for all models, showing how difficult it can be to estimate the geometry of the obscurer from X-ray spectral fitting in the low-counts regime. The top right panel shows that NGC\,3982 is predicted to be below the effective Eddington limit for dusty gas (red dotted line from \citealt{Ricci2017c}) by the majority of the models, as expected for high column densities of circum-nuclear gas. As illustrated in the bottom right panel, the inclination angle for \textsc{UXCLUMPY} was not constrained, however, we found a strong degeneracy between the inclination angle and the covering factor for \textsc{RXTorus}, since larger covering factors can accommodate smaller inclination angles whilst still being obscured. A similar trend is seen with \textsc{borus02} though \texttt{borus-coupled} with much larger uncertainties. Regarding the scattering fraction we found a large range of posterior parameter values between different models, predicting $f_{\rm scatt}$ as low as $0.9^{+3.8}_{-0.8}$\% for \textsc{pexrav} and $0.7^{+0.3}_{-0.2}$\% for \texttt{borus-decoupled} up to $7.9^{+6.2}_{-3.5}$\% estimated by \textsc{UXCLUMPY}. The bottom left panel shows a strong degeneracy between the scattering fraction and the Eddington ratio, likely due in-part from the difficulty associated with constraining the intrinsic normalisation in reflection-dominated spectra. \textsc{UXCLUMPY} shows a deviation from the other models, giving higher Thomson-scattered fractions when compared with the intrinsic accretion power on average. A plausible reason for this deviation is the treatment of the warm mirror component in our \textsc{UXCLUMPY} model setup as compared to the other models we consider. We use the physically-motivated `omni' directional warm mirror component (see Section 4.3 of \citealt{Buchner2019}) that describes warm Compton scattering from the inter-cloud volume-filling gas as well as cold Compton scattering from the dense clouds. To first order the warm mirror component with this method is a powerlaw. But, the overall Thomson-scattered flux arising from a given fraction will be lower (for a given intrinsic coronal flux) than for the simpler formalism employed in the other model setups on average.

Overall, Figure~\ref{img:degs} shows that different models can produce posterior distributions with significantly different shapes and even predict discrepant posterior parameter ranges. The corresponding posterior model {\em spectral} realizations associated with these unique posterior distributions are shown in Figure~\ref{img:realizations} to have very similar spectral shapes up to $\sim30$~keV. Above this energy, the spectral curvature of the Compton hump from each model is somewhat unique for each torus geometry but the data in this energy range is insufficient to constrain any model parameters from it. Future higher signal-to-noise broadband spectra and/or high-spectral resolution X-ray observations together with next-generation physical models will hence help to reliably distinguish between different torus geometries (see \S\ref{sec:hexp}).

\begin{figure*}[t]
    \begin{center}
    \includegraphics[width=0.7\textwidth]{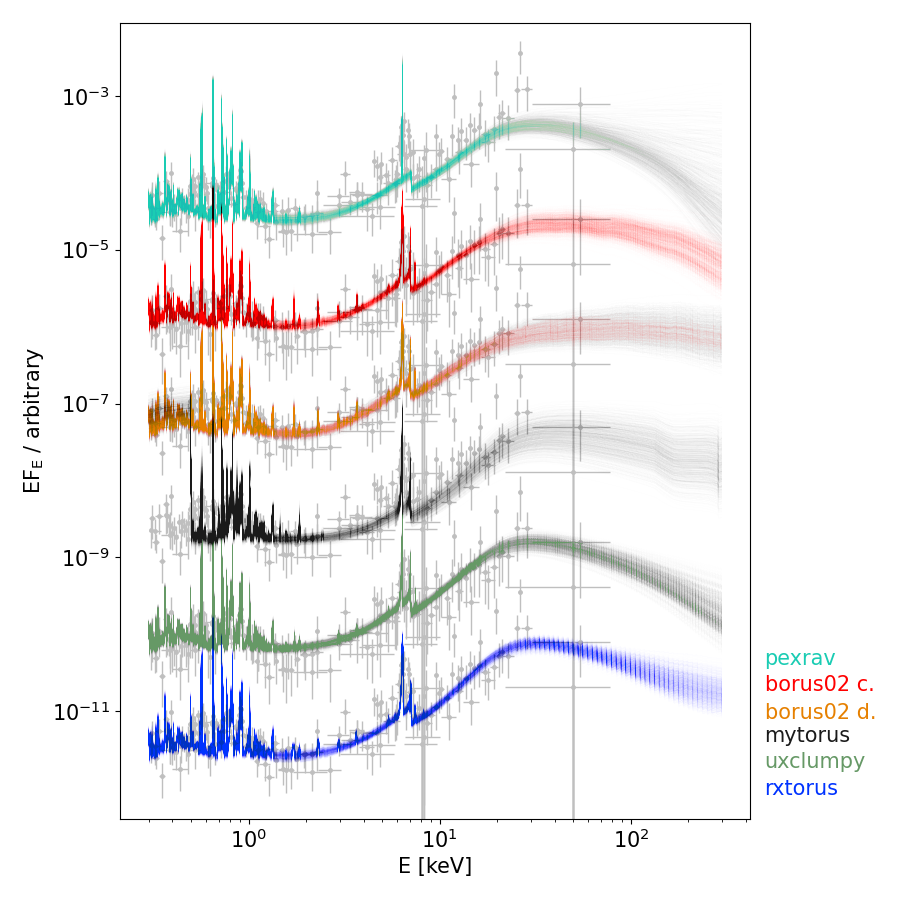}
    \caption{Posterior model realizations plotted over the spectral data unfolded with a $\Gamma=2$ power-law in grey. The overall spectral shapes predicted by each model are very similar up to $\sim30$~keV (further supported by the similar $C$/d.o.f. in Table~\ref{tab:results bxa}), with spectral differences predicted above this value. Future high-sensitivity broadband spectroscopy including $E$\,$>$\,30\,keV are hence key for distinguishing geometric models of the obscurer via the shape of the Compton hump.}
    \label{img:realizations}
    \end{center}
\end{figure*}

Studying obscured AGN down to low luminosities is important for understanding the bulk of the AGN population, as they likely form a large fraction of AGN. Many AGN population synthesis models predict that a large fraction of Compton-thick AGN is necessary to reproduce the observed spectral shape of the Cosmic X-ray Background (e.g., \citealt{Gilli2007, Treister2009, Akylas2012, Comastri2015}). However, many heavily obscured AGN harbouring a moderately accreting SMBH are beyond our capabilities for detection with current X-ray observatories, even in the local Universe at distances just above $\sim 50-100$~Mpc (e.g., \citealt{Ricci2015}). Namely NGC\,3982, would most probably remain undetected in the \textit{NuSTAR} observation if located as far as at $\sim40$~Mpc, roughly double of its Hubble distance. We also expect many heavily obscured moderately accreting SMBH at the peak of the star-formation and black hole growth at redshift $z\sim2$ (e.g. \citealp{Ueda2014}). Such sources could be significantly contributing to the cosmic black hole growth but still remain hidden behind layers of gas and dust.


\subsection{Local vs. Global parameter exploration}

To quantitatively evaluate the performance of local parameter exploration with \textsc{Xspec} in comparison with global parameter exploration from BXA, we performed Monte Carlo simulations. We loaded the corresponding model spectral realisation associated with each posterior model row in \textsc{Xspec}, as an initial position of parameter space that was by definition close to the global minimum found by BXA. We then ran Levenberg-Marquadt-based X-ray spectral fits from these starting positions many times, tracking the best-fit parameter values acquired on each iteration. Although the exact number of realizations varies from model to model, they all fall in between 2226 and 3093.

Examples for a selection of interesting parameter distribution comparisons for \textsc{MYTorus} and \textsc{UXCLUMPY} are shown in Figure~\ref{img:method comparison}, where we reiterate that the BXA posterior distribution illustrated in blue was used on average as the prior parameter guess for each \textsc{Xspec} fit used to generate the corresponding orange distributions. All orange parameter distribution shapes are significantly different from those found with BXA. Some parameter distributions are clearly consistent with the local parameter exploration algorithms getting stuck in local minima (e.g., photon index and scattered fraction), in which the mode that is consistent with BXA is significantly lower probability than a secondary mode that is entirely inconsistent with BXA. Other parameters (e.g., line-of-sight column density and inclination angle) are broadly consistent with the average distribution found with BXA, though with additional probability weight located at the extremes of the BXA distributions. The discrepancies we find from fitting with local parameter exploration imply that X-ray spectral fitting can be significantly effected by local minima, though we note that our local parameter estimates are do not consider their associated error which could lead to more overall consistency with BXA. A full quantitative comparison between global and local parameter exploration would require comprehensive simulations of different parameter values, which is outside the scope of the current work.

\begin{figure*}[t]
    \begin{center}
    \includegraphics[width=0.49\textwidth]{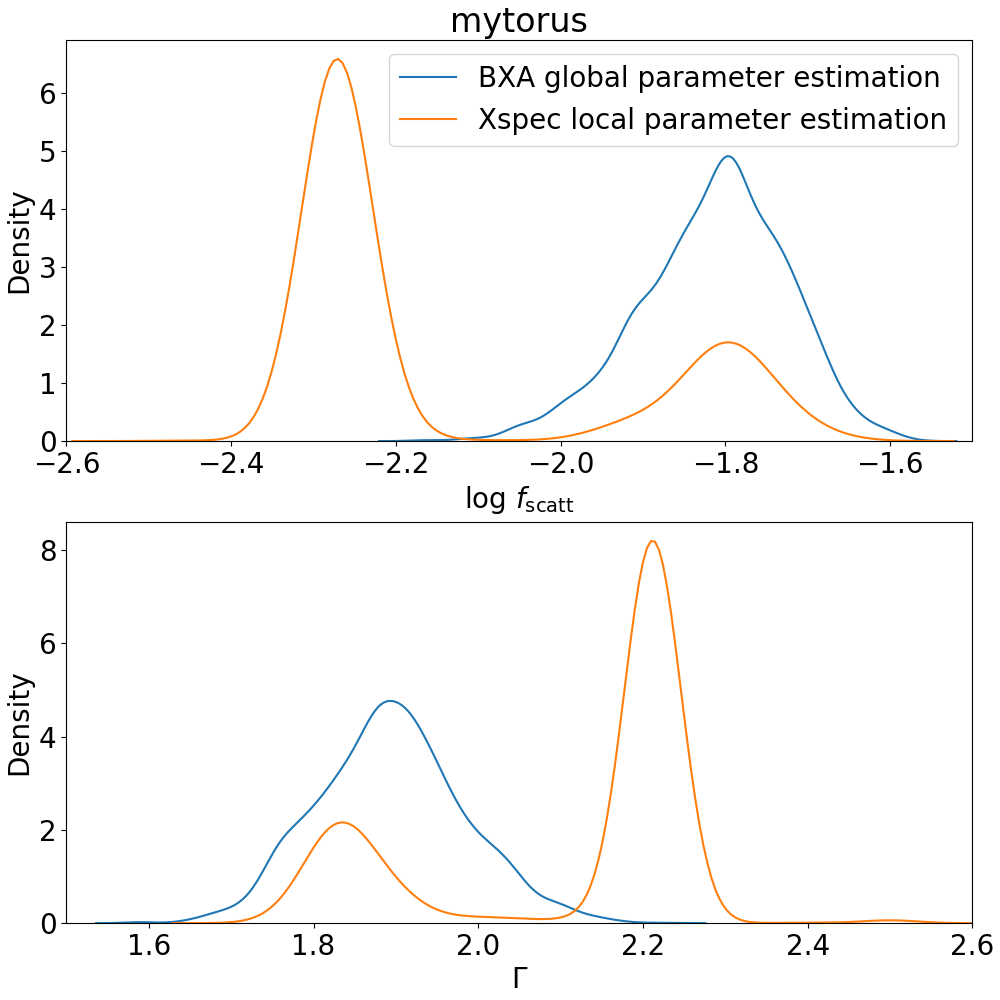}
    \includegraphics[width=0.49\textwidth]{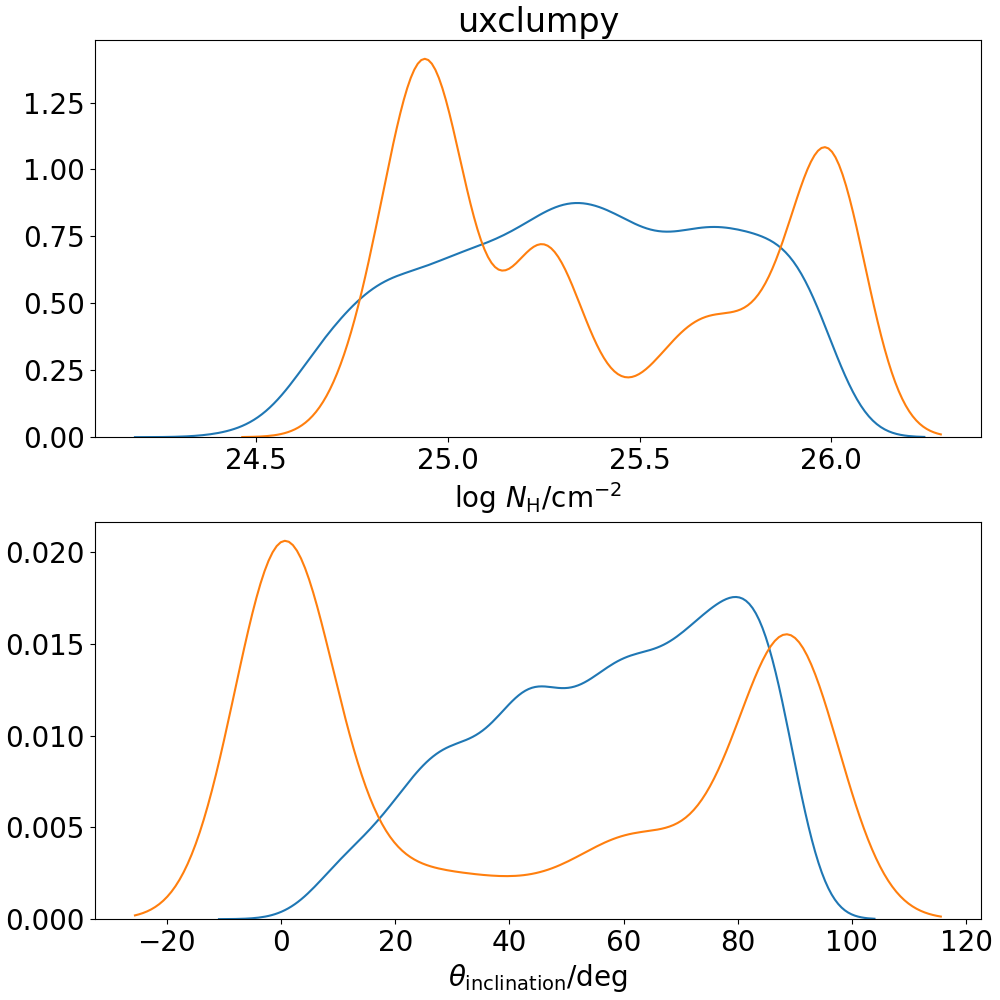}
    \caption{The posterior parameter distributions for the logarithm of the scattering fraction (log\,$f_{\rm scatt}$) and the photon index ($\Gamma$) obtained by \textsc{MYTorus} model (\textit{left} panels) and the logarithm of the line-of-sight column density (log$N_{\rm H}$) and the inclination angle ($\theta_{\rm inclination}$) for the \textsc{UXCLUMPY} model (\textit{right}) panels. The posterior values from BXA (\textit{blue}) were used to inform the initial parameter values used in the local parameter exploration with \textsc{Xspec}. The distributions of best-fit values originating from the local parameter estimation in \textsc{Xspec} are shown in \textit{orange}. All distributions shown are generated with a KDE approximation to the actual distributions, which is why for example the inclination angle distributions appear to allow negative values that are not attainable in any of the models tested.}
    \label{img:method comparison}
    \end{center}
\end{figure*}

\subsection{\textit{HEX-P} simulations}
\label{sec:hexp}

The High Energy X-ray Probe\footnote{\url{https://hexp.org}} (\textit{HEX-P}; \citealt{Madsen2023}) is a next-generation probe class mission concept that provides simultaneous broad-band coverage (0.2--80\,keV) via the combination of two hard X-ray-focusing High Energy Telescopes (HETs) and a Low Energy Telescope (LET) with significantly improved sensitivities relative to \textit{XMM-Newton} and \textit{NuSTAR} combined. Here we visualise the spectral improvements attainable with \textit{HEX-P} in studying the circum-nuclear obscurer in the bulk of the Compton-thick AGN population, by running spectral simulations from our spectral fits to NGC\,3982. We note that the \textit{HEX-P} simulations shown are conservative since NGC\,3982 is one of the lowest-luminosity Compton-thick AGN known within a volume of $\sim$20\,Mpc.

The \textit{HEX-P} response files version v07 for two HETs on-board \textit{HEX-P} were used to simulate broadband spectra, where we focused on the $2-80$~keV energy band for visualisation. The simulations were performed from our BXA \textsc{UXCLUMPY} best-fit model with nine different combinations of \texttt{TORsigma} and \texttt{CTKcover} to test a wide number of geometrically distinct configurations for the obscurer. For \texttt{TORsigma} we assumed values of 7, 28 and 84~degrees and for \texttt{CTKcover} we used 0.0, 0.3 and 0.6 with a total exposure of 200\,ks. The inclination angle was fixed to 70 degrees, while the line-of-sight column density was set to $4.2\times10^{25}~\rm cm^{-2}$. To aid comparison, we performed the same spectral simulations with the response and background files from the real \textit{NuSTAR}/FPMA observation with the same exposure as for \textit{HEX-P}.


Figure~\ref{img:hexp comparison} shows both the \textit{NuSTAR} and \textit{HEX-P} simulated spectra, normalised to unity at 7.1\,keV to show the range in Compton hump shapes expected from different obscurer geometries described by \texttt{TORsigma} and \texttt{CTKcover}. As Figure~\ref{img:hexp comparison} illustrates, \textit{HEX-P} is capable of not only constraining the spectral shape $<$\,30\,keV that is key for measuring column density and inferring intrinsic AGN parameters, but also the geometrical properties of their circum-nuclear obscurers from distinct Compton hump shapes. Such studies with \textit{HEX-P} will thus open a new era for understanding the connection between the intrinsic properties of the central source and the obscurer in heavily obscured AGN down to low intrinsic AGN powers. Such a connection is currently difficult to constrain in Compton-thick AGN, as compared to the less obscured AGN population (e.g., \citealt{Ricci2017c}).

\begin{figure*}[t]
    \makebox[\textwidth]{
    \hspace{-2.2em}
    \includegraphics[width=\textwidth]{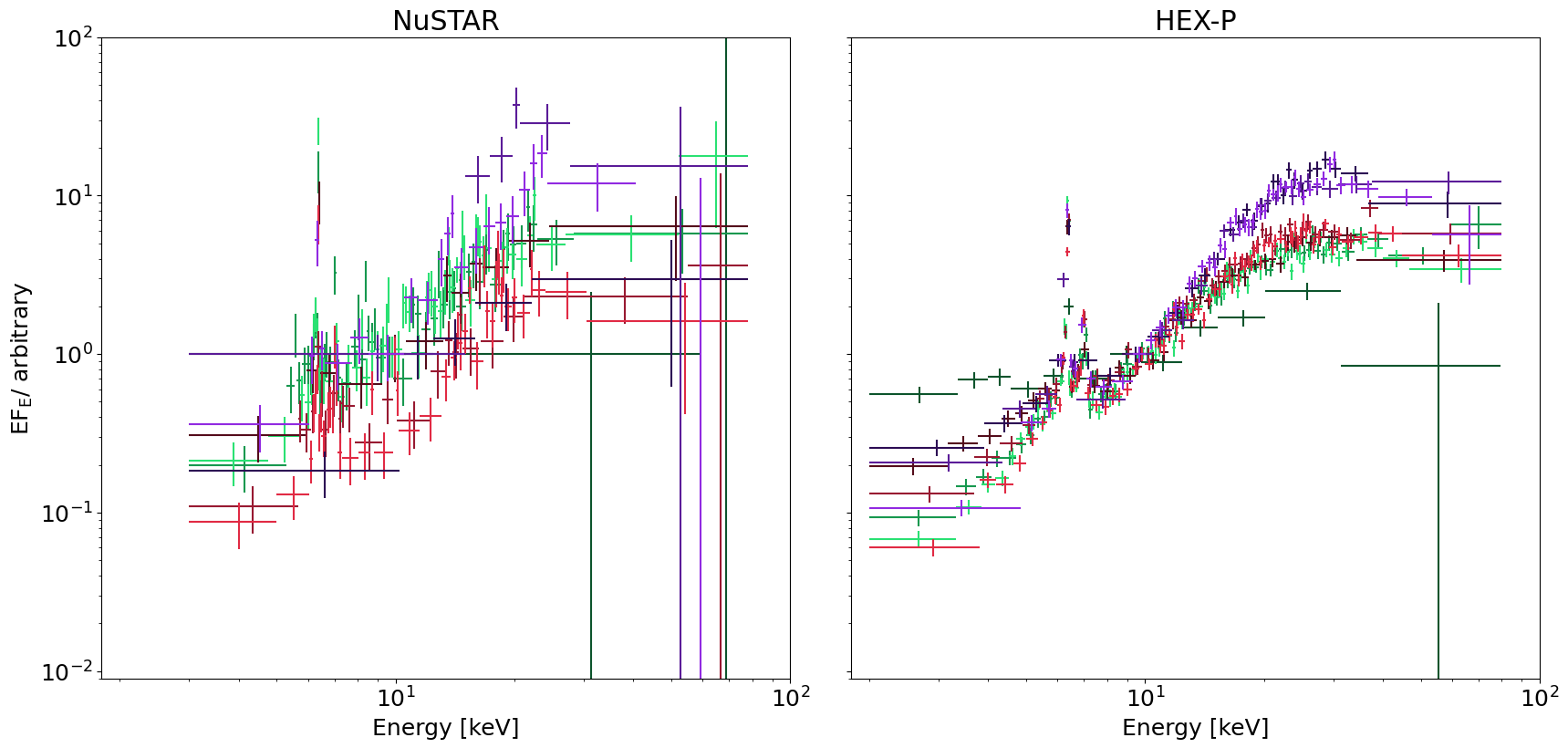}}
    \centering\includegraphics[width=0.85\textwidth]{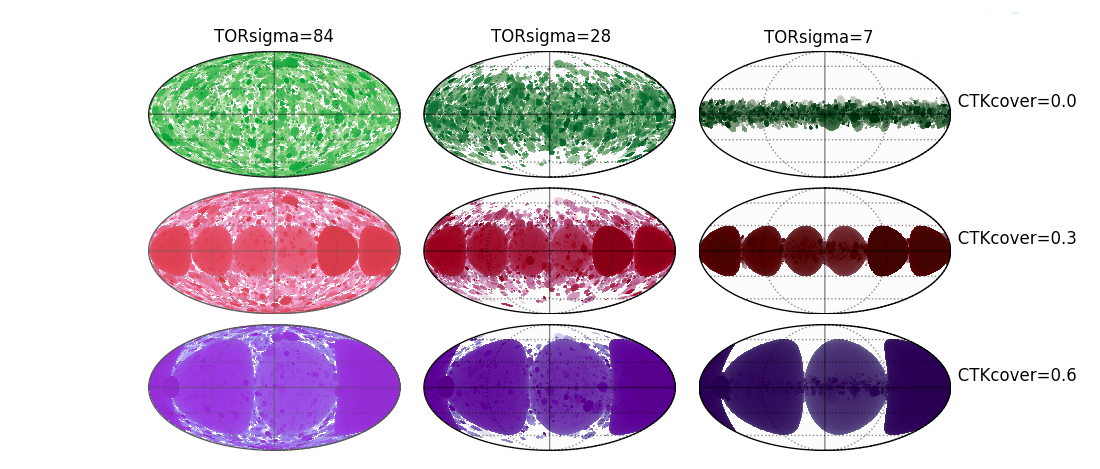}
    \caption{\textit{NuSTAR} spectra simulated using the \textsc{UXCLUMPY} model (\textit{left}) using an exposure of 200\,ks and for nine different combinations of \texttt{TORsigma} and \texttt{CTKcover}. Equivalent simulations for \textit{HEX-P} (\textit{right}) allows identification of differences in Compton hump shape arising from different obscurer geometries. Colors indicate different combinations of \texttt{TORsigma} and \texttt{CTKcover} describing the geometry of the obscurer in \textsc{UXCLUMPY}, as illustrated by the obscurer maps in the \textit{bottom} panel. The maps are adopted from the \textsc{UXCLUMPY} GitHub page\footnote{\href{https://github.com/JohannesBuchner/xars/blob/master/doc/uxclumpy.rst}{https://github.com/JohannesBuchner/xars/blob/master/doc/uxclumpy.rst}}.}
    \label{img:hexp comparison}
\end{figure*}

\section{Summary and Conclusion}
\label{sec:summary}

This work investigates the model-parameter dependencies arising from a detailed analysis of the X-ray spectral properties of the low-luminosity Sy~2 AGN NGC\,3982. We fit the broadband X-ray spectra generated from \textit{XMM-Newton} and \textit{NuSTAR} with five different obscurer models using local and global parameter exploration algorithms. Our key findings are:

\begin{itemize}
    \item \textit{Compton-thick classification:} The line-of-sight column density was found to be $>1.5\times10^{24}\,\rm cm^{-2}$ for all the models at the 3$\sigma$ confidence level, for \textsc{UXCLUMPY} reaching values as high as log\,$N_{\rm H}$/cm$^{-2}$\,=\,$24.9-25.8$ (see Table~\ref{tab:results bxa}). We thus confirm NGC\,3982 to be a Compton-thick AGN. 
    
    \item \textit{Inter-parameter dependencies:} The two-dimensional posterior parameter distributions acquired with BXA show clearly different shapes across the models considered (see Figure~\ref{img:degs}), even though some of the one-dimensional parameter distributions are comparable. We find a large range of predicted intrinsic luminosities, highlighting the difficulties associated with constraining the intrinsic continuum in reflection-dominated Compton-thick AGN.

    \item \textit{Intrinsic AGN power:} We compare the intrinsic X-ray luminosity predicted to bolometric indicators of intrinsic AGN power. The median X-ray luminosity in the $2-10$~keV band was found to be $10^{40.9-42.1}\,\rm erg\,s^{-1}$ for individual models (see Table~\ref{tab:results bxa}), which is in agreement with estimations from the MIR $12\,\mu \rm m$ emission at the 68\% confidence level for all models considered (see top left panel in Figure~\ref{img:degs}). On the other hand, we predicted a much higher X-ray luminosity from the optical [O\,III]\,$\lambda$5007 emission potentially indicating the source was more luminous in the past, in agreement with \citet{Esparza-Arredondo2020}. The Eddington ratio was found to be below the dusty-gas limit found by \cite{Ricci2017c} for all the models (see top right panel in Figure~\ref{img:degs}), with the majority of models predicting a posterior median Eddington ratio in the range $0.2-1.1$\%.

    \item \textit{Predicted spectral shapes:} Despite each model reproducing unique multi-dimensional parameter posterior distributions, we found the corresponding model realizations to have very similar spectral shapes up to $\sim30$~keV, as shown in Figure~\ref{img:realizations}. We additionally found the critical energy region for distinguishing models is $E$\,$\gtrsim$\,10\,keV, in which the current spectral constraints are insufficient.

    \item \textit{Local vs. global parameter exploration:} We perform Monte Carlo tests comparing local to global parameter exploration, finding that local parameter estimation can give clearly different results from global parameter estimation techniques for specific parameters (see Figure~\ref{img:method comparison}). This shows the advantages of algorithms such as those available with BXA in fitting complex models, in addition to the low-flux regime.

    \item \textit{HEX-P:} By simulating \textit{HEX-P} spectra of NGC\,3982 we show the benefits of simultaneous high-sensitivity broadband X-ray spectroscopy in disentangling multiple spectral components $>$10\,keV. \textit{HEX-P}, in combination with complex future torus models, will hence constrain the geometrical and physical properties of the circum-nuclear obscuring material in the low-luminosity Compton-thick AGN population as well as its connection with intrinsic AGN characteristics.

\end{itemize}

Our study highlights that X-ray spectral fitting of obscured AGN not only relies on the data being fit and the method used to perform the fitting, but inherently relies on the choice of model. Using global parameter exploration algorithms to fit multiple models is a powerful method to quantify the effects model-dependencies have on deriving key parameters from obscured AGN.

\begin{acknowledgments}
The authors are grateful to the referee for helpful comments. K.K. and C.R. acknowledge support from ANID BASAL project FB210003. P.B. acknowledges financial support from the Czech Science Foundation under Project No. 22-22643S. C.R. acknowledges support from Fondecyt Regular grant 1230345.
\end{acknowledgments}



\clearpage
\newpage
\bibliography{ngc3982}{}
\bibliographystyle{aasjournal}

\clearpage
\newpage
\appendix

\section{Posterior probability distributions}
\label{appendix:corners}

This section presents the posterior probability corner plots, giving the 1D and 2D marginalized posteriors for all the fitted parameters 
with uncertainties marked by dashed lines indicating 95\% (2$\sigma$) interval.

\begin{figure}[h]
    \makebox[\textwidth][c]{%
    \includegraphics[width=1.\textwidth]{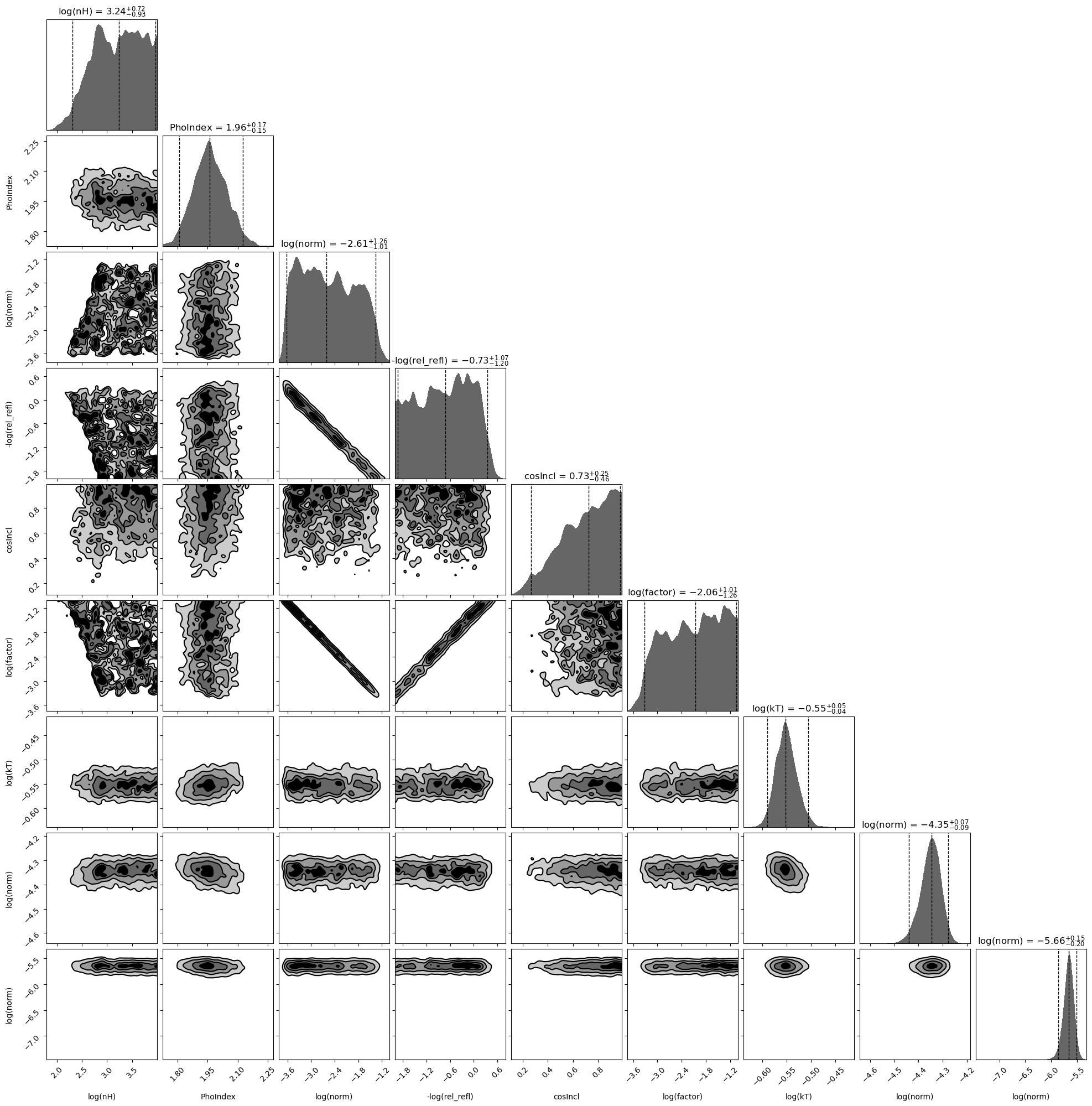}%
    }
    \caption{Posterior probability distributions for the \textsc{pexrav} model. Shown parameters in the columns (rows) are displayed as following, from left to right (top to bottom): (1) logarithm of the intrinsic column density in units of $10^{22} \ \mathrm{cm^{-2}}$; (2) photon index; (3) logarithm of the power-law normalization; (4) negative value of the logarithm of relative reflection; (5) cosine of the inclination angle; (6) logarithm of the scattering fraction; (7) logarithm of the \texttt{apec} temperature in keV; (8) logarithm of the \texttt{apec} normalization and (9) logarithm of the fluorescent Fe K$\alpha$ line normalization.}
\end{figure}

\begin{figure}[h]
    \makebox[\textwidth][c]{%
    \includegraphics[width=1.\textwidth]{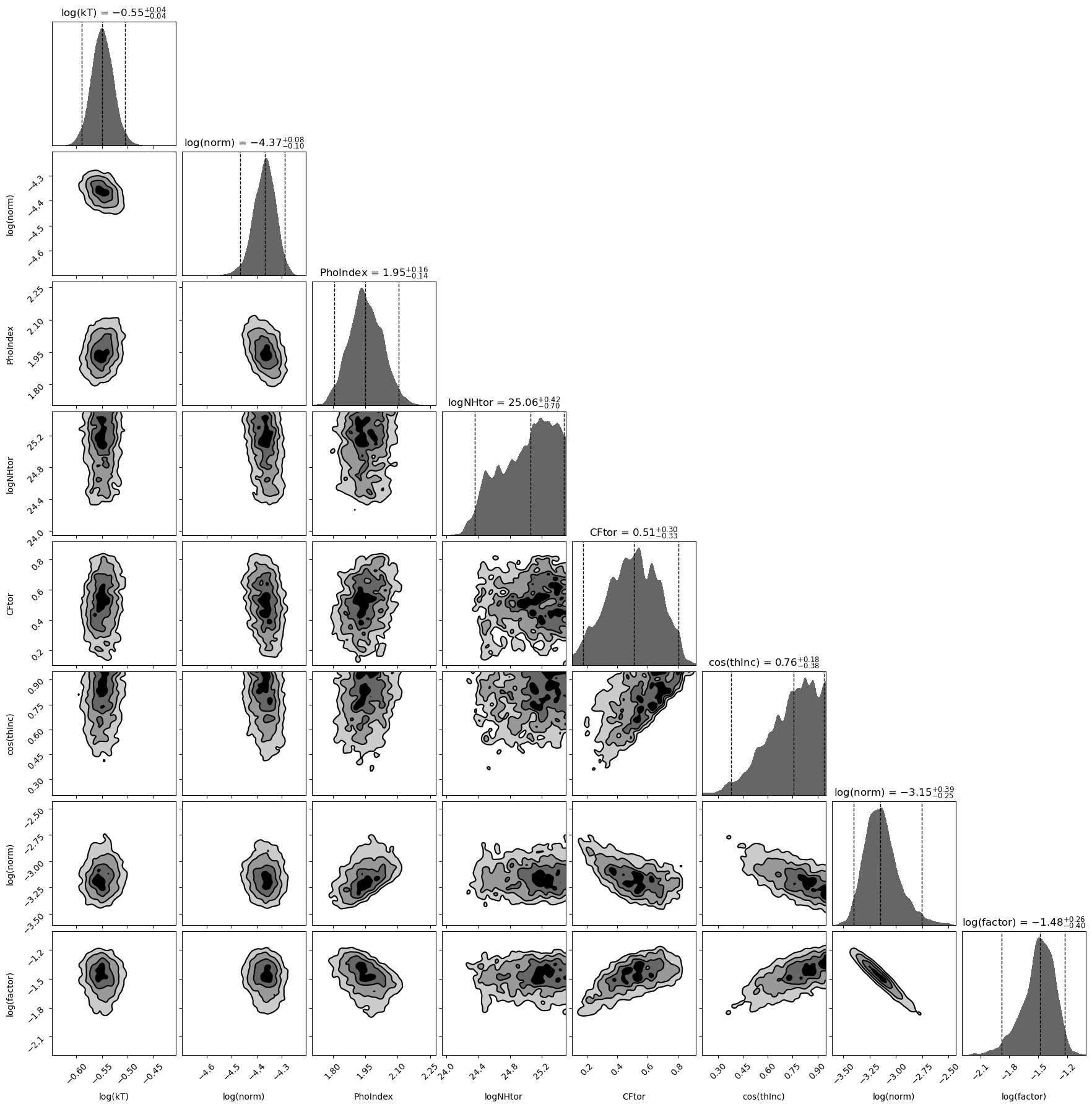}%
    }
    \caption{A corner plot showing the posterior probability distributions for the \textsc{borus02} model in the coupled mode. Columns (rows) display following parameters: (1) logarithm of the \texttt{apec} temperature in keV; (2) logarithm of the \texttt{apec} normalization; (3) photon index; (4) logarithm of the global column density of the torus; (5) covering factor of the torus; (6) cosine of the inclination angle; (7) logarithm of the power-law normalization and (8) logarithm of the scattering fraction.}
\end{figure}

\begin{figure}[h]
    \makebox[\textwidth][c]{%
    \includegraphics[width=1.\textwidth]{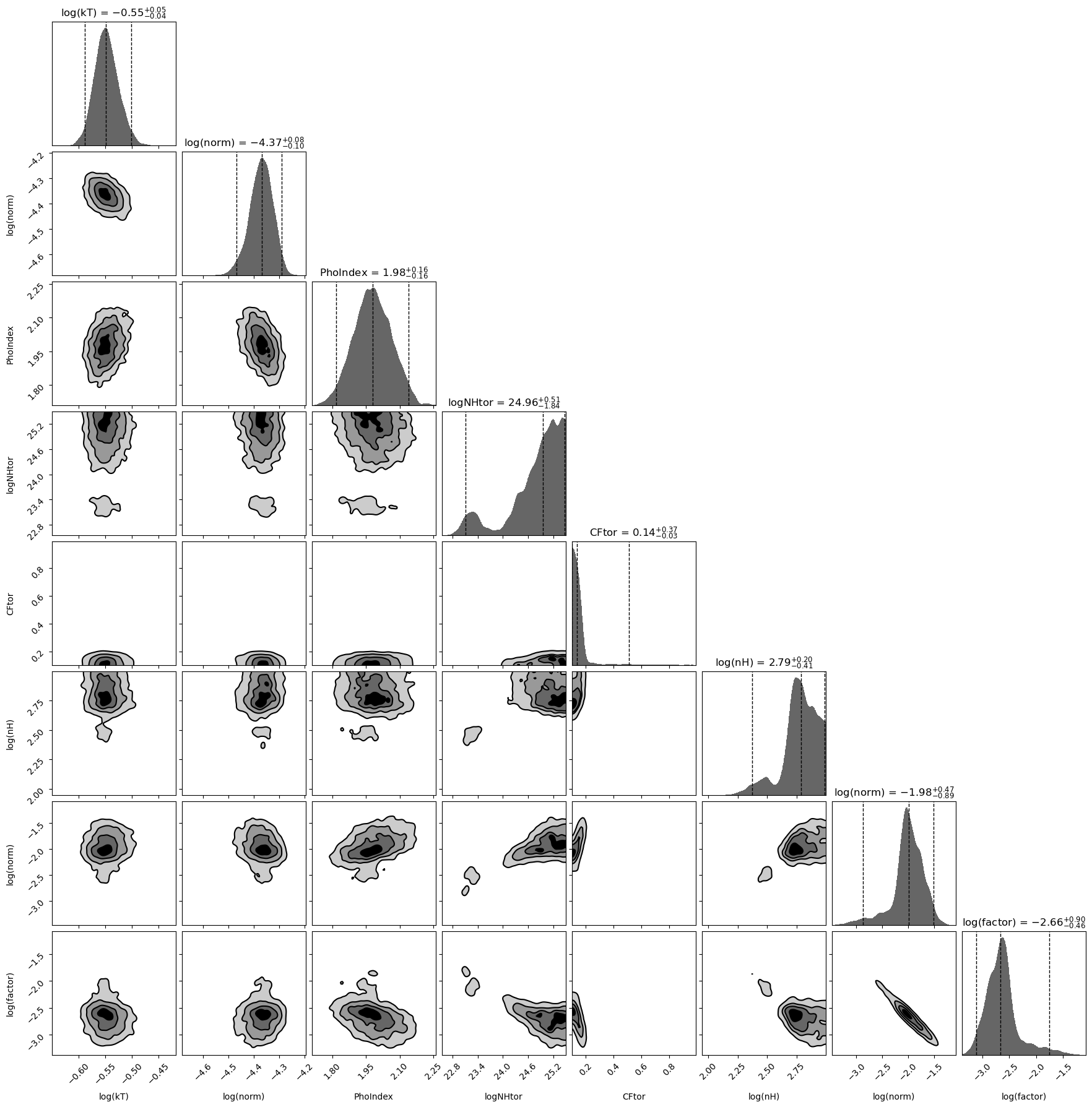}%
    }
    \caption{A corner plot showing the posterior probability distributions for the \textsc{borus02} model in decoupled mode. Columns (rows) display following parameters: (1) logarithm of the \texttt{apec} temperature in keV; (2) logarithm of the \texttt{apec} normalization; (3) photon index; (4) logarithm of the global column density of the torus; (5) covering factor of the torus; (6) logarithm of the line-of-sight column density; (7) logarithm of the power-law normalization and (8) logarithm of the scattering fraction.}
\end{figure}

\begin{figure}[h]
    \makebox[\textwidth][c]{%
    \includegraphics[width=1.\textwidth]{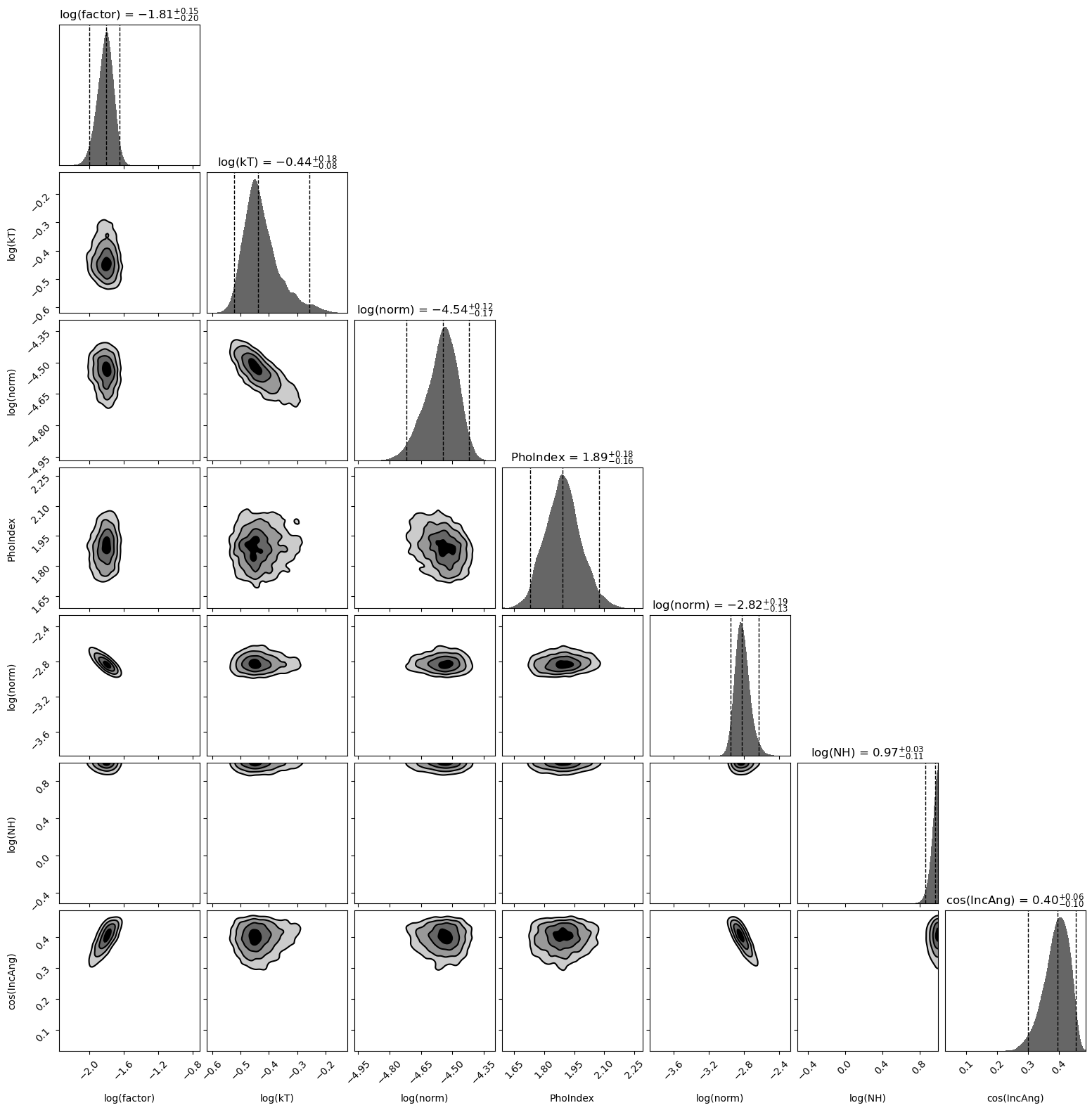}%
    }
    \caption{A corner plot showing the posterior probability distributions for the \textsc{MYTorus} model. Columns (rows) display following parameters: (1) logarithm of the scattering fraction; (2) logarithm of the \texttt{apec} temperature; (3) logarithm of the \texttt{apec} normalization; (4) photon index; (5) logarithm of the power-law normalization; (6) logarithm of the global column density in units of $10^{24} \ \mathrm{cm^{-2}}$ and (7) cosine of the inclination angle.}
\end{figure}

\begin{figure}[h]
    \makebox[\textwidth][c]{%
    \includegraphics[width=1.\textwidth]{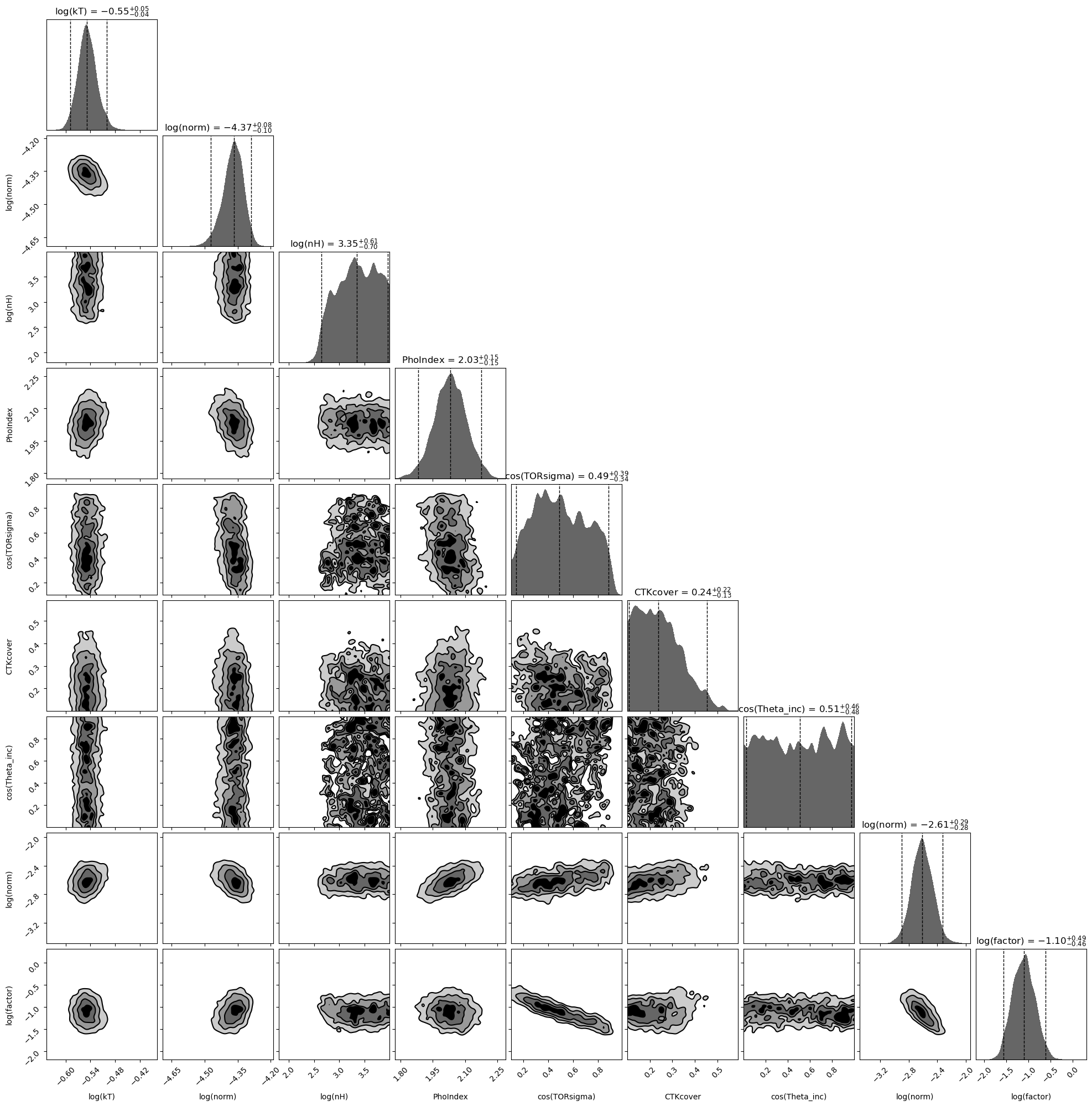}%
    }
    \caption{A corner plot showing the posterior probability distributions for the \textsc{UXCLUMPY} model. Columns (rows) display following parameters: (1) logarithm of the \texttt{apec} temperature; (2) logarithm of the \texttt{apec} normalization; (3) logarithm of the line-of-sight column density in units of $10^{24} \ \mathrm{cm^{-2}}$; (4) photon index; (5) cosine of the torus dispersion \texttt{TORsigma}; (6) Compton-thick inner ring covering factor; (7) cosine of the inclination angle; (8) logarithm of the power-law normalization and (9) logarithm of the scattering fraction.}
\end{figure}

\begin{figure}[h]
    \makebox[\textwidth][c]{%
    \includegraphics[width=1.\textwidth]{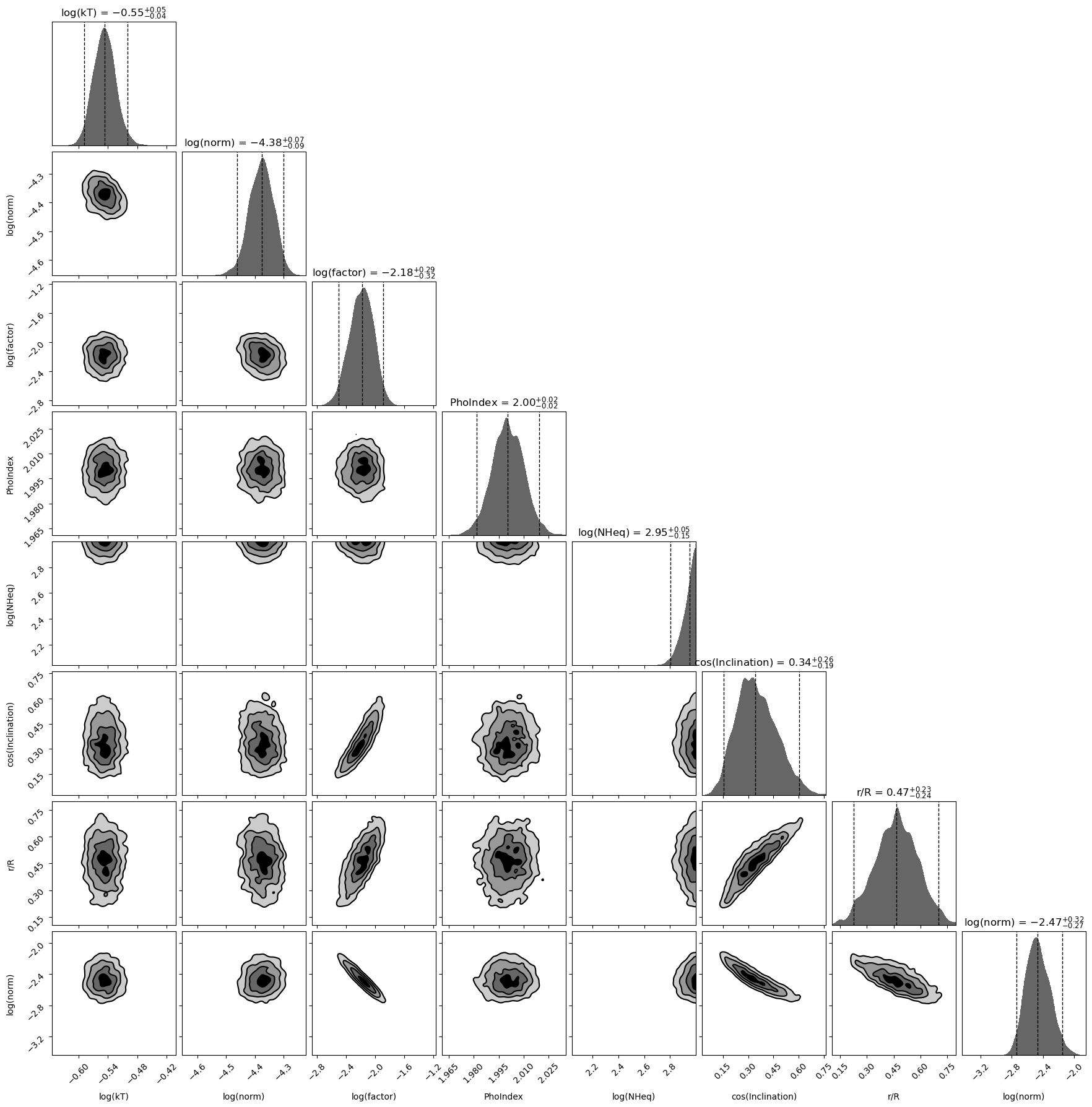}%
    }
    \caption{A corner plot showing the posterior probability distributions for the \textsc{RXTorus} model. Columns (rows) display following parameters: (1) logarithm of the \texttt{apec} temperature, (2) logarithm of the \texttt{apec} normalization; (3) logarithm of the scattering fraction; (4) photon index; (5) logarithm of the equatorial column density in units of $10^{22} \ \mathrm{cm^{-2}}$; (6) cosine of the inclination; (7) covering factor given as $r/R$ and (8) logarithm of the normalization of the power-law.}
\end{figure}

\end{document}